\documentclass[final,3p,times]{elsarticle}
\usepackage{amssymb}
\usepackage{amsmath}
\usepackage{xcolor} 
\definecolor{crossred}{RGB}{150, 35, 35}
\definecolor{checkgreen}{RGB}{15, 107, 40}

\usepackage{booktabs}
\usepackage{multirow}
\usepackage{hyperref}

\journal{Computer Methods in Applied Mechanics and Engineering}

\begin{document}
\begin{frontmatter}



\title{Graph Neural Network Surrogates for Contacting Deformable Bodies with Necessary and Sufficient Contact Detection}


\author[me]{Vijay K. Dubey} %% Author name
\author[me]{Collin E. Haese} %% Author name
\author[ase]{Osman Gültekin}
\author[glasgow]{David Dalton}
\author[ase,bme,oden]{Manuel K. Rausch} %% Author name
\author[ase,oden]{Jan Fuhg} %% Author name

\affiliation[me]{organization={Walker Department of Mechanical Engineering, The University of Texas at Austin},%Department and Organization
            city={Austin},
            postcode={78712}, 
            state={Texas},
            country={U.S.}}
            
\affiliation[ase]{organization={Department of Aerospace Engineering \& Engineering Mechanics, The University of Texas at Austin},%Department and Organization
	city={Austin},
	postcode={78712}, 
	state={Texas},
	country={U.S.}}
\affiliation[glasgow]{organization={School of Mathematics and Statistics, University of Glasgow},%Department and Organization
	city={Glasgow},
	postcode={G12 8QQ}, 
	country={U.K.}}
	
\affiliation[bme]{organization={Department of Biomedical Engineering, The University of Texas at Austin},%Department and Organization
	city={Austin},
	postcode={78712}, 
	state={Texas},
	country={U.S.}}

\affiliation[oden]{organization={The Oden Institute of Computational Science and Engineering, The University of Texas at Austin},%Department and Organization
	city={Austin},
	postcode={78712}, 
	state={Texas},
	country={U.S.}}

%% Abstract
\begin{abstract}
Surrogate models for the rapid inference of nonlinear boundary value problems in mechanics are helpful in a broad range of engineering applications. However, effective surrogate modeling of applications involving the contact of deformable bodies, especially in the context of varying geometries, is still an open issue. In particular, existing methods are confined to rigid body contact or, at best, contact between rigid and soft objects with well-defined contact planes. Furthermore, they employ contact or collision detection filters that serve as a rapid test but use only the necessary and not sufficient conditions for detection. In this work, we present a graph neural network architecture that utilizes continuous collision detection and, for the first time, incorporates sufficient conditions designed for contact between soft deformable bodies.  We test its performance on two benchmarks, including a problem in soft tissue mechanics of predicting the closed state of a bioprosthetic aortic valve. We find a regularizing effect on adding additional contact terms to the loss function, leading to better generalization of the network. These benefits hold for simple contact at similar planes and element normal angles, and complex contact at differing planes and element normal angles. We also demonstrate that the framework can handle varying reference geometries. However, such benefits come with high computational costs during training, resulting in a trade-off that may not always be favorable. We quantify the training cost and the resulting inference speedups on various hardware architectures. Importantly, our graph neural network implementation results in up to a thousand-fold speedup for our benchmark problems at inference.
\end{abstract}

\begin{graphicalabstract}
\includegraphics[width=\textwidth]{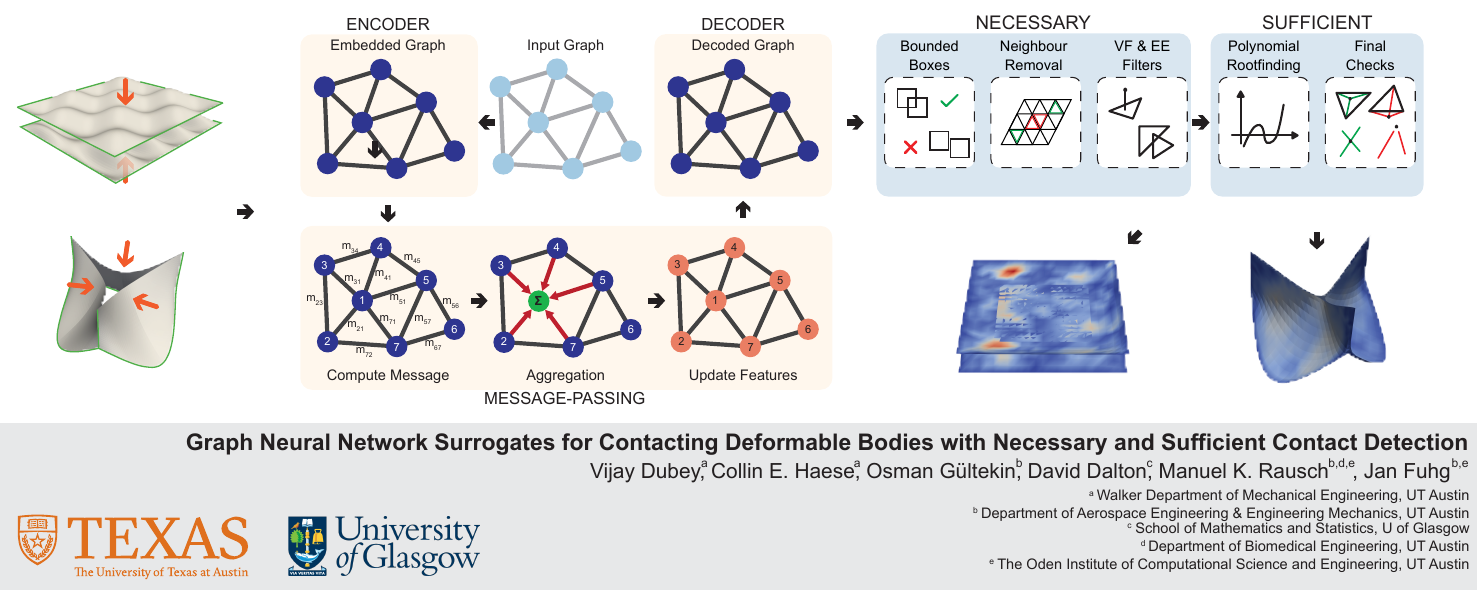}
\end{graphicalabstract}

\begin{highlights}
%\item Implementation of contact algorithm that uses both necessary and sufficient conditions for contact detection in-loop with training of a machine learning framework.
%\item Inclusion of contact penalty during data-driven training of graph neural network models regularizes the learning.
%\item Can be applied to nonlinear finite strain mechanics, accommodates varying reference configurations, and generalizes well to unseen data. 
%\item Incurs high computational cost at training time which might be reduced in future by optimizing the implementation.
%\item Provides up to a thousand fold speedup compared to a FEM solver.

\item A graph neural network with a continuous collision detection algorithm for contact penalty computation.

\item We show regularization in learning and better generalization by including this loss term.

\item Applicable to nonlinear finite-strain mechanics and accommodates varying reference configurations.

\item Provides up to a thousand-fold speedup in inference but incurs high computational costs during training.

\end{highlights}

%% Keywords

\end{frontmatter}



\section{Introduction}\label{secIntroduction}
%% Labels are used to cross-reference an item using \ref command.

The use of finite element methods (FEM) for high-fidelity modeling is ubiquitous in continuum mechanics. With decades of development, the FEM can handle both geometric and material non-linearities \cite{zienkiewicz_general_2014}, and incorporate multiple contacting bodies \cite{wriggers_computational_2006}. However, their iterative solution process incurs high computational and time costs that often grow super-linearly with the size of the problem. This computational cost is a bottleneck in the use of high-fidelity FEM models in applications that require near-real-time predictions. Surrogate modeling can help alleviate this issue \cite{fuhg2021state} through its ability to learn and accelerate simulations, whether in solid or fluid mechanics, and has shown varying degrees of success in the areas of computational solid mechanics \cite{aldakheel_what_2022}. However, due to its complexity, effective surrogate modeling of applications involving contact of deformable bodies, especially in the context of varying geometries, is still an open issue \cite{goodbrake2024neural}.

% Talk about contact being important but most work has been done for non-contacting bodies

% Talk about intrusive methods (ROM) then go over that non-intrusive methods based on ML can be more flexible because they can work with black box solvers and allow to parameterize geometries (which is really important in contact)

%The computational costs of FEM models further increase significantly when contact between bodies are present and has to be modeled. This stems primarly from two reasons: (i) finer meshes, and (ii) smaller time steps. 

Generally, we can distinguish between intrusive and non-intrusive surrogate models  \cite{padula2024brief}. Intrusive methods directly modify or access the internal structure of the high-fidelity model, i.e., the partial differential equation (PDE), while non-intrusive methods define a surrogate without having access to the full-order operators or to their action on a vector \cite{ghattas2021learning}. Both archetypes have recently been employed for surrogate modeling of contact mechanics problems. However, since contact is a complex problem driven by PDEs under constraints, surrogate modeling remains an area of active research.

\textbf{Intrusive Surrogate Models for Contact.} \, 
The most commonly employed intrusive surrogate modeling technique in computational mechanics is based on using singular value decomposition (SVD) on snapshot data for generating a reduced-order basis \cite{amsallem2012nonlinear}. To the best of the authors' knowledge, this approach was used for the first time on contact mechanics problems by Krenciszek et al. \cite{krenciszek2014model}. Balajewicz et al. \cite{balajewicz_projection-based_2016}  propose to use different approaches to build the primal reduced basis and the dual reduced basis. In particular, the primal reduced basis is obtained through standard SVD, while the dual reduced basis is built using nonnegative matrix factorization (NNMF) on the Lagrange multiplier snapshots. Thereby, the method guarantees positive basis vectors.
 This method is extended further using hyper-reduction techniques such as the discrete version of the empirical interpolation method (DEIM) by Fauque, et al., \cite{fauque_hybrid_2018} and others \cite{benaceur2020reduced, pawar2025efficient}. 
 However, \cite{kollepara2023limitations} have pointed out that low-rank model order reduction techniques, which are effective for many physical systems, struggle to accurately represent the contact pressure field in contact mechanics due to its inherently high-dimensional and spatially localized nature. They show that the dual space remains high-dimensional, even under aggressive truncation.
 
% Problems of intrusive modeling for contact

Recently, physics-informed neural network models have also been employed as intrusive surrogates for contact mechanics. These weakly enforce the physical laws, evaluated at a set of collocation points, in the loss function. They have been used in the context of energy minimizations, see \cite{goodbrake2024neural}, or by employing the strong form of the PDE \cite{sahin2024solving,sahin2024physics}.
%Overall, while intrusive surrogate modeling has proven to successfully speed up selected problems in contact mechanics, they are still not fully equipped to deal with the inherent non-smoothness, inequality constraints, and complementarity conditions present in contact mechanics problems, particularly in the presence of friction, contact of multiple deformable bodies, or self-contact. 
Intrusive surrogate modeling has successfully accelerated certain problems in contact mechanics. However, it is still not fully equipped to address some inherent challenges. These challenges include non-smoothness, inequality constraints, and complementarity conditions, especially when friction, multiple deformable bodies, or self-contact are involved.
These complicate reduced formulations and require direct access to full solvers, including robust contact search algorithms.

\textbf{Non-intrusive Surrogate Models for Contact.} \, 
Non-intrusive surrogate models rely on access to a dataset from a high-fidelity solver. They focus on minimizing data fit errors against the ground-truth values. Using this concept, non-intrusive surrogate models for selected parts of the contact response (such as surface topography to friction coefficient maps) between deformable bodies have been developed in recent years that use feedforward neural networks \cite{eskinazi2015surrogate,kalliorinne2021artificial,houssein2025artificial}, or Gaussian process regression \cite{sahin2025data}. 
%For data-driven methods relying on ground truth generated using the mesh-based solution, the data can naturally be represented as graphs.
Graph-based surrogates are ideal for data-driven methods relying on ground truth generated using the mesh-based solution due to the natural representation of meshes as graphs.
Since contact mechanics problems are commonly solved using the finite element method, meshed geometries and their deformations are usually available from the solver. Therefore, there is an obvious link to surrogate models based on graph representations.
 Graph representations allow for relational inductive biases that are invariant on the node and edge permutation \cite{hamilton_graph_2020}\cite{battaglia_relational_2018}. Furthermore, they can learn arbitrary relations, which makes them ideal for dynamical predictions in complex systems \cite{battaglia_interaction_2016, sanchez-gonzalez_learning_2020, mrowca_flexible_2018, belbute-peres_combining_2020}. Similar graph frameworks have been adapted for both data-driven and physics-informed approaches for cardiac mechanics \cite{dalton_emulation_2022}\cite{dalton_graph_2021}.  
 
In the context of contacting bodies, one of the biggest challenges for graph neural networks (GNNs) is the consistent detection and prevention of contact penetration.
GNNs have been applied for the deformation of a hyper-elastic plate deformed by a rigid actuator and flag dynamics with additional specialized structure to learn relations based on proximity in physical space, which the authors use to handle collision or contact \cite{pfaff_learning_2020}. This approach which the authors termed MeshGraphNets relies on changes to the architecture and not additional loss terms. FIGNet \cite{allen_learning_2022} is a similar approach that recognizes the complexity of collision detection and adds representative face-face edge features to the structure, based solely on the bounding volume hierarchy (BVH). Others have used fast contact filters to regularize data-driven training for a case of rigid tool and deformable body contact \cite{zhu_collision-aware_2022}; these filters utilize the necessary but not sufficient conditions for contact or collision. This approach builds on past work that suggests improved learning outcomes by adding specialized physics-based loss terms \cite{lee_data-driven_2019}\cite{wang_towards_2020}. 

\vspace{0.5cm}

However, to the best of the authors' knowledge, no studies to date have implemented a full-fledged contact algorithm (bounding volumes and necessary as well as sufficient conditions) to regularize graph-based data-driven learning for contact between soft bodies. The current work is the first to present such a framework and to discuss its implications. Table \ref{table-lit-compare} compares the current work with similar or related works in the past.

\begin{table}[htb]
\centering
\resizebox{\textwidth}{!}{
\begin{tabular}{cccccccccc}
\toprule
 &
  \begin{tabular}[c]{@{}c@{}}Mesh\\ Space \\ Edge\end{tabular} &
  \begin{tabular}[c]{@{}c@{}}World\\ Space\\ Edge\end{tabular} &
  \begin{tabular}[c]{@{}c@{}}Contact \\ between\\ Soft Bodies\end{tabular} &
  \begin{tabular}[c]{@{}c@{}}Contact\\ Detection\\ Filter\end{tabular} &
  \begin{tabular}[c]{@{}c@{}}Precise\\ Contact\\ Detection*\end{tabular} &
  \begin{tabular}[c]{@{}c@{}}Contact\\ Loss\\ Term\end{tabular} &
  \begin{tabular}[c]{@{}c@{}}Code\\ Availability\end{tabular} &
  \begin{tabular}[c]{@{}c@{}}Dataset\\ Availability\end{tabular} &
  \begin{tabular}[c]{@{}c@{}}Hierarchical\\ Graph\end{tabular} \\ \midrule
Dalton, et al. \cite{dalton_emulation_2022}     & \textcolor{checkgreen}{\checkmark} &  \textcolor{crossred}{X} & \# & \textcolor{crossred}{X}  & \textcolor{crossred}{X}  & \textcolor{crossred}{X}  &  \textcolor{checkgreen}{\checkmark} &  \textcolor{checkgreen}{\checkmark} &  \textcolor{checkgreen}{\checkmark}\\
Pfaff, et al. \cite{pfaff_learning_2020}     & \textcolor{checkgreen}{\checkmark} &  \textcolor{checkgreen}{\checkmark} & \textcolor{crossred}{X} & \textcolor{crossred}{X}\ddag  & \textcolor{crossred}{X}  & \textcolor{crossred}{X}  &  \textcolor{checkgreen}{\checkmark} &  \textcolor{checkgreen}{\checkmark} & \textcolor{crossred}{X}\\
Allen, et al. \cite{allen_learning_2022}     & \textcolor{checkgreen}{\checkmark} &  \dag & \# & \textcolor{crossred}{X}\ddag  & \textcolor{crossred}{X}  & \textcolor{crossred}{X}  &  \textcolor{checkgreen}{\checkmark} &  \textcolor{checkgreen}{\checkmark} & \textcolor{crossred}{X}\\
Zhu, et al. \cite{zhu_collision-aware_2022}      &  \textcolor{checkgreen}{\checkmark} &  \textcolor{checkgreen}{\checkmark} & \textcolor{crossred}{X}  &  \textcolor{checkgreen}{\checkmark} & \textcolor{crossred}{X}  &  \textcolor{checkgreen}{\checkmark} & \textcolor{crossred}{X}  &  \textcolor{checkgreen}{\checkmark} & \textcolor{crossred}{X}\\
This work &  \textcolor{checkgreen}{\checkmark} &  \textcolor{checkgreen}{\checkmark} &  \textcolor{checkgreen}{\checkmark} &  \textcolor{checkgreen}{\checkmark} &  \textcolor{checkgreen}{\checkmark} &  \textcolor{checkgreen}{\checkmark} &  \textcolor{checkgreen}{\checkmark} &  \textcolor{checkgreen}{\checkmark} & \textcolor{crossred}{X}\\ \bottomrule
\end{tabular}
}
\caption{Comparison of the current framework with existing frameworks. \dag: FIGNet generalizes the notion of world-space edge from nodes to faces as well. \#: Works suggest potential application to contact between soft bodies. \ddag: However, these utilize pairwise distance and collision radius to build edge relations. *: Both necessary and sufficient conditions for contact or collision detection.}
\label{table-lit-compare}
\end{table}

 In the current work, we extend a graph neural network framework that has continuous collision detection to incorporate sufficient conditions for contact. This collision detection is used to compute a penalty term representative of the extent of the contact violation. Due to its continuous nature, it is robust enough to handle cases where the contact constraint violation occurs at any time between a given time step (and not just at the end) and cases where the elements completely go through each other during a time step. This algorithm is general enough to allow contact between multiple soft deformable bodies, unlike others in the past who have applied it to rigid-soft body interactions \cite{zhu_collision-aware_2022} where the contact interface is known in advance or to rigid-rigid body interactions \cite{allen_learning_2022}. 
 We are aware that this entails a significant increase in training time compared to existing methods. However, from a computational contact mechanics point of view, this is a natural step forward. Our intention is to provide a proof-of-concept for the community and to gauge whether this additional training time is necessary. 
 
 We apply the framework to benchmark problems and show that the developed algorithm regularizes the training, leading to better generalization. Furthermore, we show that these aspects are true even for a complex contact case wherein the contact planes move and the element normal angles are not uniform. Additionally, we demonstrate more than a thousand-fold speedup in inference time. To provide a complete picture, we provide an inference speedup on multiple hardware architectures and training time for a single epoch on one of these platforms. In the interest of open science, we share our complete datasets and code base on the Texas Data Repository and Github, respectively.

The work is arranged as follows: first, Section \ref{secMethods} describes the framework. This includes the architecture of the graph neural network (Section \ref{sec-architecture}), the implemented contact algorithm (Section \ref{sec-collisiondetection}), the type of losses used (Section \ref{sec-lossfunctions}), and additional details on the implementation and the training process (Section \ref{sec-implementation-and-training}). Second, we describe the two benchmark problem datasets used to test the framework (Section \ref{sec-benchmarkproblems}). Finally, we define performance metrics (Section \ref{sec-performancemetrics}), present the results based on these metrics (Section \ref{sec-results-discussion}) and discuss them. 

\section{Methods}\label{secMethods}
%% Use \subsubsection, \paragraph, \subparagraph commands to 
%% start 3rd, 4th and 5th level sections.
%% Refer following link for more details.
%% https://en.wikibooks.org/wiki/LaTeX/Document_Structure#Sectioning_commands

\subsection{Architecture}\label{sec-architecture}

\subsubsection{Graphs and Notation}\label{sec-GraphsAndNotation}
In the following, we use uppercase and lowercase to refer to graph and vertex indices, respectively. We use Graph Neural Networks (GNNs) that operate on an input graph $(G_I)$. A graph consists of objects that form its vertices $(x)$ and their relationships, which form the edges $(e)$. In general, depending on the object(s) and relationship(s) chosen, there can be more than one vertex type and/or edge type. The information is organized at the vertex-level, edge-level, and graph-level. We denote a vertex-feature using $x_{Ii}$ for the i-th vertex in the I-th graph. Similarly, there is an edge-feature $e_{Iij}^k$ for a k-th type of edge between vertices $(i,j)$ in the I-th graph. Finally, we denote the graph-level feature by $g_I$ for the I-th graph. The different types of edges are differentiated using their superscripts. 

\subsubsection{Features}\label{sec-Features}
The ground truth data arising from the solution of a finite element simulation can be represented as graphs. Each time step in the solution data is one graph, each node is a vertex, and each element edge is an edge in a graph. Like Pfaff et al. \cite{pfaff_learning_2020}, we define, in addition to existing `mesh-space’ edges, an edge-type based on positional proximity of nodes in the physical space referred to as `world-space' connectivity. We use a similar notation to theirs and refer to these `mesh-space' and `world-space' edges using superscripts $M$ and $W$ respectively. 
\paragraph{Graph Feature $(g_I)$} Simulation parameters that are fixed for all vertices of a graph form its graph features. In our case, this is the absolute time in the solution $t$, the time step size $\delta t$, the material constants, and other information such as applied pressure. %For example, the problem described in section \ref{secApplicationAV} has graph feature $g_I = [t, \delta t, c_0, c_1, c_2, p]$ where $p$ is the pressure and $c_i$ are the material constants.
\paragraph{Node Features $(x_{Ii}, y_{Ii})$} The state of each node consists of its current position $r_{Ii}$, current velocity $v_{Ii}$ and current acceleration $a_{Ii}$. This together forms input nodal features $x_{Ii} = [r_{Ii},v_{Ii},a_{Ii}]$ at current time $t$. The goal is to predict the nodal accelerations at time $t+\delta t$ denoted by $y_{Ii}$.
\paragraph{Edge Features $(e_{Iij}^M, e_{Ikl}^W)$} The mesh-space edge is defined only between vertices that are connected by a finite element mesh, irrespective of their position; it is denoted by $e_{Iij}^M$ (see Equation \eqref{eq-def-meshspaceedge}). Similarly, we have a world-space edge denoted by $e_{Ikl}^W$ between all nodes closer than a collision radius $R$ (see Equation \eqref{eq-def-worldspaceedge}):
\begin{align}
e_{Iij}^M &= \left[r_{Ii}-r_{Ij}, \|r_{Ii}-r_{Ij}\|_{L^2}\right] \forall (i,j) \in \text{ element connectivity} \label{eq-def-meshspaceedge}, \\
e_{Ikl}^W &= \left[r_{Ik}-r_{Il}, \|r_{Ik}-r_{Il}\|_{L^2}\right] \forall (k,l) \text{ such that } \|r_{Ik}-r_{Il}\|_{L^2} \leq R .\label{eq-def-worldspaceedge}
\end{align}

The graph $G_I$ consists of all its graph features, input-nodal features, and edge features. Thus, $G_I = (g_I, x_{Ii}, e_{Iij}^M, e_{Ikl}^W)$ where the indices run as stated previously. For each graph, we have a set of ground truth target values $(y_{Ii})$.

\begin{figure}[htb]%% placement specifier
	%% Use \includegraphics command to insert graphic files. Place graphics files in 
	%% working directory.
	\centering%% For centre alignment of image.
	\includegraphics[width=\textwidth]{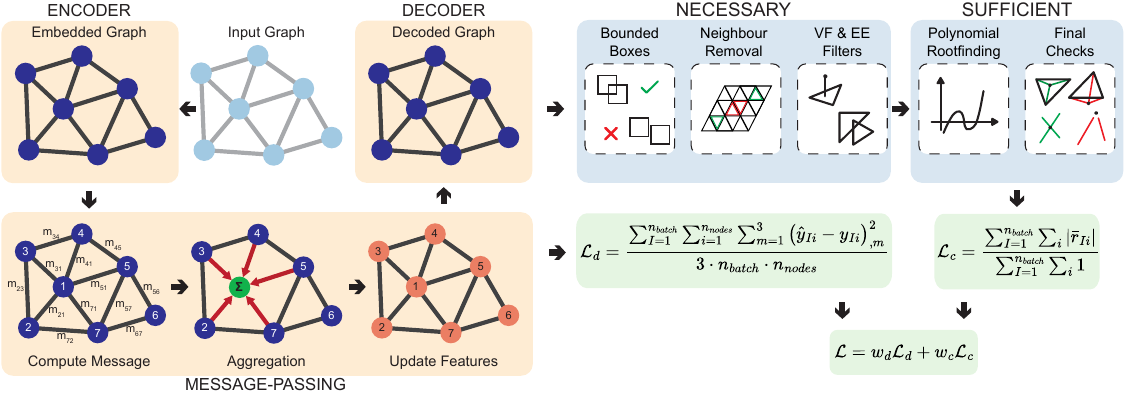}
	%% Use \caption command for figure caption and label.
	\caption{Schematic of the framework used in this work.\textbf{Orange boxes} are core components of the graph neural network. \textbf{Blue boxes} highlight the steps of the contact algorithm as described in the Section \ref{sec-collisiondetection}. This is split into steps for necessary conditions of collision followed by sufficient conditions of collision. \textbf{Green boxes} indicate the computed losses which are combined using weighted sum.}\label{fig_Framework}
	%% https://en.wikibooks.org/wiki/LaTeX/Importing_Graphics#Importing_external_graphics
\end{figure}
\subsubsection{Neural Network} \label{sec-neuralnetwork}
We utilize an encode-process-decode architecture similar to those previously used in \cite{dalton_emulation_2022}\cite{pfaff_learning_2020} with the message passing processor. Each part in this architecture utilizes a standard fully connected feed-forward neural network (simply MLPs hereon) with a nonlinear activation function. The rectified linear unit is used as an activation function. Skip connections are present in the hidden layers of the MLPs.
\paragraph{Encoders} There are four types of encoders (see Equation \eqref{eq-encoders}) each of which embeds a corresponding feature from the input graph $(g_I, x_{Ii}, e_{Iij}^M, e_{Ikl}^W)$ to the corresponding encoded feature, denoted here by boldface version of the same symbol ${(\mathbf{g}_I, \mathbf{x}_{Ii}, \mathbf{e}_{Iij}^M, \mathbf{e}_{Ikl}^W)}$. Each of the encoders is an MLP denoted by $\mathcal{E}_g$, $\mathcal{E}_x$, $\mathcal{E}_e^M$, and $ \mathcal{E}_e^W$ respectively:
\begin{align}
\mathbf{g}_I = \mathcal{E}_g(g_I) ;\quad \mathbf{x}_{Ii} = \mathcal{E}_x (x_{Ii}) ;\quad \mathbf{e}_{Iij}^M = \mathcal{E}_e^M(e_{Iij}^M) ;\quad \mathbf{e}_{Iij}^W = \mathcal{E}_e^W(e_{Iij}^W). \label{eq-encoders}
\end{align}

\paragraph{Message Passing (Processor)} The encoded nodal and edge features are used for message passing. They consist of three steps: message creation, aggregation, and update. The message creation is described in Equation \eqref{eq-msggen}. A distinct message is created for each type of edge. In particular, on the mesh-space edge denoted by ${m^M_{Iij}}$ and on the world-space edge denoted by ${m^W_{Ikl}}$. At the cost of additional computation, we enforce skew-symmetry of the generated messages to strengthen their interpretation as forces between the nodes \cite{dalton_emulation_2022} (see Equation \eqref{eq-msgskew}). For aggregation, we define a one-hop neighborhood of node $Ii$ in the mesh-space denoted by $\mathcal{N}^M(Ii)$ (and $\mathcal{N}^W(Ii)$ in world-space) and aggregate all messages from edges within this neighborhood of the central node. The aggregation operator $\bigoplus$ is the sum of all messages that arrive at the central node. Finally, the aggregations from the mesh-space and world-space edges are used to update the embedded feature of the central node. The aggregation and update step is described in Equation \eqref{eq-msgpass-aggupdate}. Finally, the generated messages are used to update the embedded edge-features (Equation \eqref{eq-msgpas-edgeupdate}). The message generation function $\phi$ and the update function $\gamma$ are both MLPs. The message passing process (Equations \eqref{eq-msggen} to \eqref{eq-msgpas-edgeupdate}) can be repeated using distinct generation and update functions in each round, referred to by the index $n$ for up to $k$ rounds. Therefore, the process reads:
\begin{align}
m^{M(n)}_{Iij} = \phi^{M(n)} \left(\mathbf{x}_{iI}^{(n)}, \mathbf{x}_{Ij}^{(n)},\mathbf{e}_{Iij}^{M(n)}\right) \quad&;\quad m^{W(n)}_{Iij} = \phi^{W(n)} \left(\mathbf{x}_{iI}^{(n)}, \mathbf{x}_{Ij}^{(n)},\mathbf{e}_{Iij}^{W(n)}\right), \label{eq-msggen}\\
\tilde{m}^{M(n)}_{Iij} = \frac{m^{M(n)}_{Iij} - m^{M(n)}_{Iji}}{2} \quad&;\quad \tilde{m}^{W(n)}_{Iij} = \frac{m^{W(n)}_{Iij} - m^{W(n)}_{Iji}}{2}, \label{eq-msgskew}\\
\mathbf{x}_{Ii}^{(n+1)} = \gamma^{(n)} &\left( \mathbf{x}_{Ii}^{(n)} , \bigoplus_{j \in \mathcal{N}^{M}(Ii)} \tilde{m}^{M(n)}_{Iij} , \bigoplus_{j \in \mathcal{N}^{W}(Ii)} \tilde{m}^{W(n)}_{Iij}  \right), \label{eq-msgpass-aggupdate} \\
\mathbf{e}_{Iij}^{M(n+1)} = \mathbf{e}_{Iij}^{M(n)} + \tilde{m}^{M(n)}_{Iij} \quad&;\quad \mathbf{e}_{Iij}^{W(n+1)} = \mathbf{e}_{Iij}^{W(n)} + \tilde{m}^{W(n)}_{Iij}. \label{eq-msgpas-edgeupdate}
\end{align}

\paragraph{Decoder} The dot product of the encoded graph-level features $\mathbf{g}_I$ with embedded nodal features $\mathbf{x}_{Ii}^{(k+1)}$ after $k$ rounds of message passing is used for decoding.  Three different decoders ($\mathcal{D}_i; i\in\{1,2,3\}$), each of which is an MLP, are used to get each component of the nodal acceleration as shown in Equation \eqref{eq-decoders}: 
\begin{align}
\mathcal{X}_{Ii} &= \mathbf{x}_{Ii}^{(k+1)} \cdot \mathbf{g}_I \label{eq-graph-dot-nodalfeature}\, ,\\
\hat{y}_{Ii} &= [\mathcal{D}_1(\mathcal{X}_{Ii}), \mathcal{D}_2(\mathcal{X}_{Ii}), \mathcal{D}_3(\mathcal{X}_{Ii})] \, .\label{eq-decoders}
\end{align}

\subsection{Contact Algorithm}\label{sec-collisiondetection}

MeshGraphNets \cite{pfaff_learning_2020} take advantage of the relational inductive biases of graph neural networks \cite{battaglia_relational_2018} and incorporate world-space edges as a mechanism to establish a relationship between nodes that are close in physical space. This adds structure within the framework for learning relations arising from physical proximity, such as collision or contact. Others \cite{zhu_collision-aware_2022} have utilized additional contact terms to regularize learning for the specialized case of tool-object interactions. In this work, we take this one step further to address the more general case of contact between deformable bodies. 

The framework developed so far predicts the nodal accelerations, which are then integrated twice to compute the position for each node at the next time point. A continuous collision detection algorithm is used to identify each colliding pair of triangles along their trajectory in the time step $(t, t+\delta t)$. The uniform average velocity is used for the trajectory of each vertex between the time points. 

First, axes aligned bounding boxes are used to cull non-colliding pairs of triangles. A check is implemented to explicitly remove neighboring triangles connected through mesh-space to reduce false positives. Second, vertex-face and edge-edge filters are used to identify all triangle pairs that are coplanar at any time during the trajectory in the given time step \cite{zhu_collision-aware_2022}\cite{tang_fast_2010}, a necessary but not sufficient condition for collision. Only triangle pairs deemed to have become coplanar are processed in further steps. 

\begin{figure}[htb]%% placement specifier
	%% Use \includegraphics command to insert graphic files. Place graphics files in 
	%% working directory.
	\centering%% For centre alignment of image.
	\includegraphics[width=\textwidth]{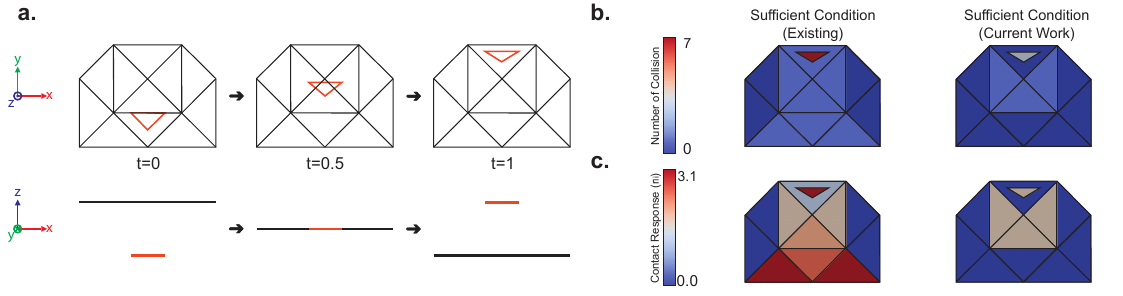}
	%% Use \caption command for figure caption and label.
	\caption{Changes due to modification of edge-edge sufficiency condition present in \cite{hansmann_collision_1997}. \textbf{(a)} Top and front view of two meshes, in black and orange, initially separated by a distance that move across each other. Snapshots at initial time, time of collision, and final time. \textbf{(b)} The number of collisions detected for each triangle. Note without the modification the smaller mesh incorrectly detects false positives. \textbf{(c)} The contact response (eq. \eqref{eq-define-maxmaxcontactresponse}) for each triangle. Note previously detected false positives now return zero contact response.}\label{fig_contact_necessary_suff}
\end{figure}

Unlike \cite{zhu_collision-aware_2022}, a sufficient condition for collision is implemented using the procedure described in \cite{hansmann_collision_1997}. For each vertex-face or edge-edge pair, a cubic polynomial is formed, the roots of which are the exact time points where the coplanarity occurs. Newton's method is used to find each root of these polynomials. All roots outside of the time step are discarded. For those within the time step: (i) for each vertex-face pair, a system of linear inequalities is solved to determine whether the vertex lies within the face, and (ii) for each edge-edge pair, a check is implemented to determine whether the two line segments in three-dimensional space intersect with each other. The latter of the two conditions replaces equation (2) of \cite{hansmann_collision_1997} to avoid false positives where collinear edges are detected as colliding edges, illustrated in Figure \ref{fig_contact_necessary_suff}.

\begin{figure}[htb]%% placement specifier
	%% Use \includegraphics command to insert graphic files. Place graphics files in 
	%% working directory.
	\centering%% For centre alignment of image.
	\includegraphics[width=\textwidth]{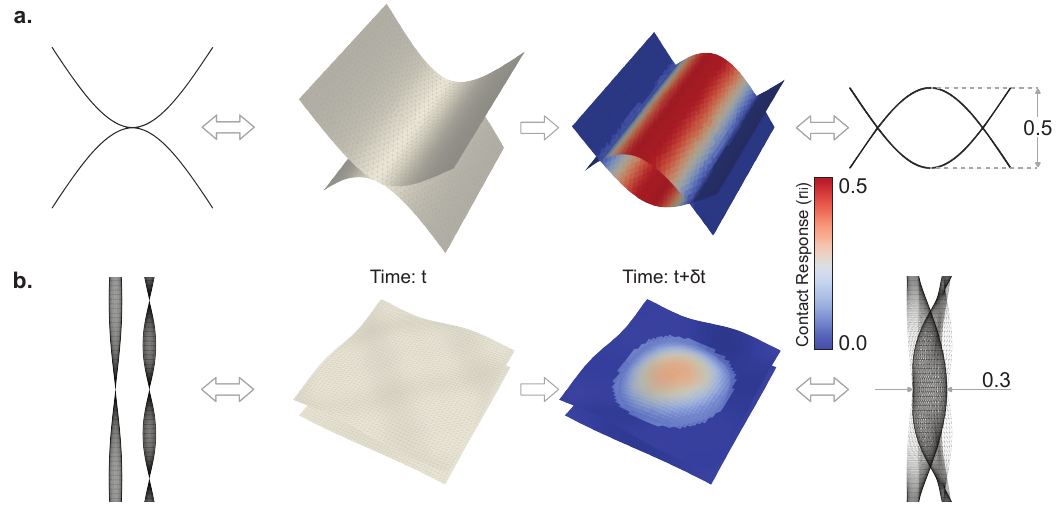}
	%% Use \caption command for figure caption and label.
	\caption{Computed contact response (Equation \eqref{eq-define-maxmaxcontactresponse}) using the implemented contact algorithm. \textbf{a.} Two parabola-like meshes move toward each other across a given time step. Some elements pass each other completely during the step. \textbf{b.} Two rectangular meshes with imposed undulations (Equation \eqref{eq-Bench2Surface}) have fixed boundaries and inflate towards each other. Unlike \textbf{a}, the contact is asymmetric. As in \textbf{a.}, some elements move past each other completely.}\label{fig_contactresponse}
	%% https://en.wikibooks.org/wiki/LaTeX/Importing_Graphics#Importing_external_graphics
\end{figure}

After ascertaining that a pair of triangles collide during a time step using the collision detection algorithm, we compute the contact response. Each triangle pair $(i,j)$ has six vertex-face and nine edge-edge responses. We use the maximum response value among these (Equation \eqref{eq-define-maxcontactresponse}). Furthermore, since a single triangle can potentially collide with multiple triangles, we take the maximum response across all such colliding pairs for each triangle (Equation \eqref{eq-define-maxmaxcontactresponse}). This computed contact response for two test cases is shown in Figure \ref{fig_contactresponse}. The key strength of the algorithm is that it is a continuous collision detection. It identifies collisions even for cases where the elements go past each other completely during a time step. This strength is observable in regions of peak contact response values in both Figure \ref{fig_contactresponse}.a and \ref{fig_contactresponse}.b. The contact response is zero when two triangles do not contact or collide during a time step.
\begin{align}
r_{Iij-VF}^p &=  
\begin{cases} 
& \text{orthogonal distance for the $p^{th}$ vertex-face test evaluated at time } (t+\delta t), \\
& \text{0 otherwise.}
\end{cases}\\ \label{eq-contact-response}
r_{Iij-EE}^q &=  
\begin{cases} 
& \text{distance between mid-points of two edges for the $q^{th}$ edge-edge test evaluated at time } (t+\delta t), \\
& \text{0 otherwise.}
\end{cases}\\
r_{Iij} &= \max_{\forall p,q} \left\{r_{Iij-VF}^p , r_{Iij-EE}^q\right\}, \label{eq-define-maxcontactresponse} \\
r_{Ii} &= \max_{\forall j} r_{ij}.\label{eq-define-maxmaxcontactresponse}
\end{align}

We give special attention to optimizing the algorithm for computational efficiency and memory usage. The collision pairs and the response are stored in sparse COO format tensors. The algorithm is vectorized with as few iterators as possible. Batch dimensions are added at specific places to establish an upper bound on memory requirements. For root finding, we exploit the structure of cubic polynomials \cite{yuksel_fast_2022}. Newton iterations are implemented, not as a while loop conditional on convergence, but rather as for loops wherein the loop body has been lifted to a compiled function. The compilation (via \texttt{torch.compile}) has been used to wrap methods where such usage does not lead to incorrect behavior. Finally, Horner's scheme is utilized to evaluate polynomials whenever necessary.

\subsection{Loss Functions}\label{sec-lossfunctions}

\paragraph{\textbf{Dynamic Loss}} Given a set of graphs $\mathcal{S}$ (in a specific batch), where each graph is defined as in Sections \ref{sec-GraphsAndNotation} \& \ref{sec-Features}, the \textit{dynamic loss} $\mathcal{L}_d$ is defined as the mean square error in the predicted nodal accelerations (Equation \eqref{eq-decoders}) and the ground-truth nodal acceleration ($y_{Ii}$) across all components, each node $[1, n_{nodes}]$, and all graphs $[1,n_{batch}]$ in $\mathcal{S}$ (Equation \eqref{eq-loss-dynamic}):
\begin{align}
\mathcal{L}_d &= \frac{\sum_{I=1}^{n_{batch}}\sum_{i=1}^{n_{nodes}}\sum_{m=1}^3{\left( \hat{y}_{Ii} - y_{Ii} \right)_{,m}^2}}{ 3\cdot  n_{batch}\cdot n_{nodes}}, \label{eq-loss-dynamic} 
\end{align}
where  $(\cdot)_{,k}$ refers to the k-th component of vector $(\cdot)$.
\paragraph{\textbf{Contact Loss}} The contact response (Equation \eqref{eq-define-maxmaxcontactresponse}) is normalized using the length-scale of the problem $(l_c)$ (Equation \eqref{eq-normalizedresponse}). The mean absolute error of this normalized response is the unsupervised term which defines the contact-loss $\mathcal{L}_c $ (Equation \eqref{eq-define-contactloss}):
\begin{align}
 \bar{r}_{Ii} &= \frac{r_{Ii}}{l_c}, \label{eq-normalizedresponse} \\
 \mathcal{L}_c &= \frac{\sum_{I=1}^{n_{batch}}\sum_{i} |\bar{r}_{Ii}|}{\sum_{I=1}^{n_{batch}}\sum_{i} 1} \, .\label{eq-define-contactloss}
\end{align}
\paragraph{\textbf{Total Loss}} The total loss $\mathcal{L}$ is computed as the weighted sum of the dynamic loss (Equation \eqref{eq-loss-dynamic}) and contact loss (Equation \eqref{eq-define-contactloss}):
\begin{align}
\mathcal{L} = w_d \mathcal{L}_d + w_c \mathcal{L}_c \label{eq-totalloss}.
\end{align}

\subsection{Implementation and Training Process}\label{sec-implementation-and-training}
The implementation is done in PyTorch \cite{paszke2017automatic} \& PyTorch Geometric \cite{Fey_Lenssen_2019} and relies on the Adam optimizer \cite{kingma_adam_2014}. Multiple networks are trained with just dynamic loss or both dynamic and contact loss included (see Section \ref{sec-res-networks-training} for more details). All datasets are split in the ratio of 8:1:1 into training, validation, and test sets.
We utilize a step function for the learning rate schedule. The matrix multiplication approach is disabled for \texttt{torch.cdist()} to favor a slower but more precise distance computation for proximity calculations for building world-space edges.

\section{Numerical Experiments}\label{sec-numerical-experiments}

\subsection{Benchmark Problems}\label{sec-benchmarkproblems}
\begin{figure}[htb]%% placement specifier
	%% Use \includegraphics command to insert graphic files. Place graphics files in 
	%% working directory.
	\centering%% For centre alignment of image.
	\includegraphics[width=\textwidth]{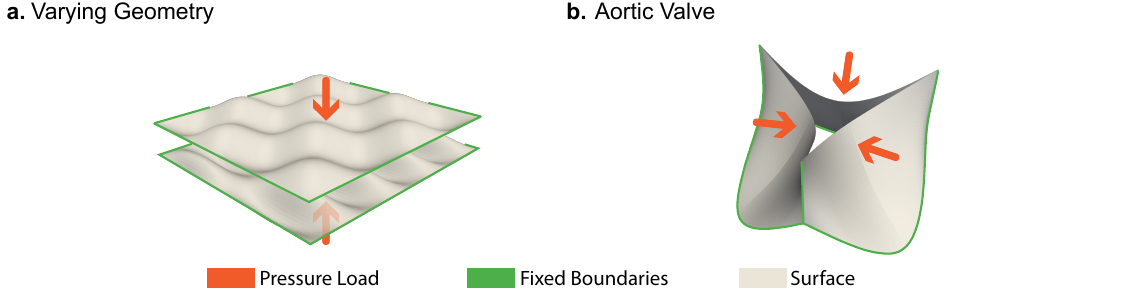}
	%% Use \caption command for figure caption and label.
	\caption{Problems used for the numerical experiments. \textbf{(a)} Varying geometry membrane wherein material properties, pressure load, and geometry is varied, and \textbf{(b)} bioprosthetic aortic valve.}\label{fig_problems}
	%% https://en.wikibooks.org/wiki/LaTeX/Importing_Graphics#Importing_external_graphics
\end{figure}

We apply the presented framework to two problems of different graph sizes, length scales, dataset sizes, and contact complexities (see Figure \ref{fig_problems} and Table \ref{tab-problem-compare}). For each problem, contact occurs between deformable soft bodies. The dataset is generated through finite element simulations using FEBio \cite{maas_febio_2012}. 

\subsubsection{Problem 1: Inflating Membranes with Varying Reference Geometry}\label{secBench2}

A gap separates two rectangular sheets with fixed edges that have a pressure load applied to their outer faces, pushing them towards each other as shown in Figure \ref{fig_problems}(a). The sheets have undulations in the reference configuration described by:
\begin{equation}\label{eq-Bench2Surface}
\begin{aligned}
        z_{[u/l]} &= A\sin{k_{1[u/l]} x}\sin{k_{2[u/l]} y} + c_3  \, , \\
            k_{1[u/l]},k_{2[u/l]} &\in \{\pi/2, \pi, 2\pi, 4\pi\} \text{ where (u,l) are (upper, lower) surfaces}.\
\end{aligned}
\end{equation}
The constant $c_3$ is chosen to be half the distance between the mean plane of each sheet and the amplitude $A$ to be smaller than half the gap distance. While the amplitude for undulation for each sheet is the same, the $\{k_1,k_2\}$ can be different for the upper and lower membranes (denoted by subscripts $(k_{xu}, k_{xl})$ respectively) and across the simulations. The material is isotropic linear elastic. This problem dataset consists of 25 simulations in which Young's modulus and the applied pressure load are sampled from a uniform random distribution. Note that, unlike the next benchmark problem, the contact is asymmetric, i.e., (a) the contact does not need to occur at the mid-plane, and (b) the contact occurs at a wide range of element normal angles. This is necessary to clearly establish the framework's ability to learn `contact' wherever it occurs instead of `learning to stop at the mid-plane'.

\subsubsection{Problem 2: Bioprosthetic Aortic Valve}\label{secApplicationAV}

We build a bioprosthetic aortic valve model (Figure \ref{fig_problems}(c)) geometry using the procedure described by \cite{xu_framework_2018}. The valve thickness and diameter are taken from \cite{xu_framework_2018} and \cite{thubrikar_aortic_2017}, respectively. The belly-edge is the only free edge, while other edges are fixed. The material is an isotropic linear-exponential model given by the strain energy function $\Psi$:
\begin{equation}
\begin{aligned}
\Psi &= \frac{c_0}{2} \left(I_1 - 3\right) + \frac{c_1}{2} \left(e^{c_2 \left(I_1 - 3\right)^2} - 1\right) \label{eq-isotropiclinearexp},
\end{aligned}
\end{equation}
where $I_1$ is the first invariant of the Cauchy-Green deformation tensor. We use the baseline values provided in \cite{wu_anisotropic_2018}. The mean blood pressure is set at 95 mmHg \cite{standring_grays_2016}. This dataset contains 25 simulations. To create the data, we vary the material parameters $(c_0,c_1,c_2)$ and the applied pressure load $P$. This case represents a simpler case of contact where the surfaces meet at known planes and the element angles are symmetric about this plane. However, the model needs to find and detect the contact itself. 

\begin{table}[!hbt]
\centering
\normalsize
\begin{tabular}{@{}ccccc@{}}
\toprule
                           & \textbf{Symbol}          & \textbf{Varying Geometry} & \textbf{Aortic Valve} \\ \midrule
Contact Plane Symmetry     & $-$  & No & Yes \\
Same Contact Plane Location & $-$ & No & Yes \\
Num. of Simulations        & $n_s$        & 25       & 25       \\
time steps per Simulation  & $n_t$        & 41       & 21       \\
Num. of Graphs             & $n_s\cdot n_t$  & 1025 & 525 \\
Num. of Nodes              & $n_n,n_{nodes}$       & 1250     & 1014     \\
Num. of Elements           & $n_{face} $      & 2304     & 1872     \\
Num. of Mesh-space Edges   & $n_e$    & 7104     & 5772     \\
Num. of Global Parameters  & $n_g$  & 2 & 4 \\
time step Size              & $\delta t$      & 0.025    & 0.025    \\
Collision Radius           & $R$ & $1.45\times10^{-1}$ &  $6.32\times10^{-4}$\\
Length Scale               & $l_c$& 2.84 & $1.26\times10^{-2}$ \\ \bottomrule
\end{tabular}
\caption{Details of the used datasets.}
\label{tab-problem-compare}
\end{table}

\subsection{Generated Datasets}
The varying geometry (Section \ref{secBench2}) and aortic valve (Section \ref{secApplicationAV}) problems have respectively six $(E, p, k_{1u}, $ $k_{2u}, k_{1b}, k_{2b})$ and four $(c_0, c_1, c_2, p)$ parameters that vary across the simulation. We add absolute time in the simulation $(t)$ and the time step $(\Delta t)$ to form graph-level features for all graphs in the dataset. For the varying geometry example, we exclude the shape parameters $( k_{1[u/l]},  k_{2[u/l]})$ to test the network's ability to learn complex and differing reference geometries for similar problems. Note that the time step size, time in simulation, and the shape parameters are discrete variables, while the rest are continuous variables. Uniform random sampling is done to generate datasets. \ref{app-datasetdetails} describes the value of parameters and their distribution for each problem in detail.

\section{Results \& Discussion}\label{sec-results-discussion}

\subsection{Performance Metrics}\label{sec-performancemetrics}
The nodal positions are computed for the FEM and for the framework developed herein using Equation \eqref{eq-GNN-FEM-process}. The mean absolute error of the displacement step prediction $(\Delta \hat{r}_{Ni})$ against the ground truth displacement step $(\Delta r_{Ni})$ normalized by the problem length scale $(l_c)$ serves as position loss $(\mathcal{L}_p)$ (Equation\eqref{eq-position-loss}). Unlike the previously discussed losses, the position loss is directly interpretable, and its value on the training and validation sets is monitored during the training process. During training, we compute position loss over batches and monitor the average of all batches in the epoch, while in inference, we do so over each graph and report the statistics. Similarly, we monitor the contact loss (Equation \eqref{eq-define-contactloss}) and report statistics on each graph after inference. We also report the quartile boundaries of the prediction error set $E$ defined in Equation \eqref{eq-error-set}. Finally, we accumulate the one-step prediction errors across all time points of a simulation (Equations \eqref{eq-accposerr}, \eqref{eq-avgaccposerr}) and report their evolution.

\begin{equation}
\begin{aligned}
r_{(N+1)i} &=\text{FEM}(r_{Ni}) & & & & & & & \hat{r}_{(N+1)i} &=\text{GNN}(r_{Ni})\\ \label{eq-GNN-FEM-process}
\Delta r _{Ni} &= r_{(N+1)i} - r_{Ni} & &\rightarrow & \xi_{Ni}&= \Delta r_{Ni} - \Delta \hat{r}_{Ni} & &\leftarrow & \Delta \hat{r}_{Ni} &= \hat{r}_{(N+1)i} - r_{Ni} \\ 
%\mathcal{R}_{Mi} &= \sum_{N=0}^{M-1} \Delta r _{Ni} + r_{0i} & &\rightarrow  & \sum_{N=0}^{M-1}\xi_{Ni} &= \mathcal{R}_{Mi} - \hat{\mathcal{R}}_{Mi} & &\leftarrow & \hat{\mathcal{R}}_{Mi} &= \sum_{N=0}^{M-1} \Delta \hat{r}_{Ni} + r_{0i}\\ 
\end{aligned}
\end{equation}

% Removed the metric 2b line
%  &\text{\textbf{Metric 2b:} } & &E = \left\{e_{Ni} | i\in[1,n_{nodes}] \text{ \& } N \in [1, n_{graphs}]\right\}  & &\text{where }e_{Ni} = \frac{\|\Delta \hat{r}_{(N+1)i} - \Delta r_{(N+1)i}\|_{L_2}}{\Sigma_i\|{\Delta r_{(N+1)i}}\|_{L_2}/\Sigma_i{1}}  \\
% Removed the metric 3 line because we need more intuitive result for paper we call metric 5: which is the new line
% &\text{\textbf{Metric 3:} } & &\Xi_{Mi} = \bigl\|\frac{\sum_{N=0}^{M-1}\xi_{Ni}}{l_c}\bigr\|_{L_2} = \bigl\|\frac{\hat{\mathcal{R}}_{Mi} - \mathcal{R}_{Mi}}{l_c}\bigr\|_{L_2} \quad \text{where } M \geq 1 ;\quad \Xi_{M} =1 \frac{\sum_{i=1}^{n_{nodes}} \Xi_{Mi} }{\sum_{i=1}^{n_{nodes}} 1}\label{eq-metric3}
\begin{align}
&\text{\textbf{Position Loss:} } & &\mathcal{L}_p = \frac{\sum_{I=1}^{n_{batch}}\sum_{i=1}^{n_{nodes}}\sum_{m=1}^3{|\left( \Delta \hat{r}_{Ni} - \Delta r _{Ni} \right)_{,m}|}}{   n_{batch}\cdot n_{nodes}\cdot 3 \cdot l_c}, \label{eq-position-loss} \\
&\text{\textbf{Position Error Set:} } & &E = \left\{e_{Ni} \mid e_{Ni} = \| \xi_{Ni}/ l_c\|_{L_2}, i\in[1,n_{nodes}] \text{ \& } N \in [1, n_{graphs}]\right\},   \label{eq-error-set} \\
&\text{\textbf{Average Position Errors:} } & &\bar{e}_N = \frac{\sum_{i=1}^{n_{nodes}} e_{Ni}}{n_{nodes}}\quad\text{;}\quad \bar{e} = \frac{\sum_{N=1}^{n_{graphs}}\sum_{i=1}^{n_{nodes}} e_{Ni}}{n_{graphs}\cdot n_{nodes}},\label{eq-avgposerr} \\
&\text{\textbf{Accumulated Position Error:} } & &\Xi_{Mi} = \sum_{N\leq M-1}e_{Ni} \quad \text{where } M \geq 1 ;\quad \Xi_{M} =1 \frac{\sum_{i=1}^{n_{nodes}} \Xi_{Mi} }{\sum_{i=1}^{n_{nodes}} 1}, \label{eq-accposerr} \\
&\text{\textbf{Average Acc. Position Errors:} } & &\bar{\Xi}_N = \frac{\sum_{i=1}^{n_{nodes}} \Xi_{Ni}}{n_{nodes}}\quad\text{;}\quad \bar{\Xi} = \frac{\sum_{N=1}^{n_{graphs}}\sum_{i=1}^{n_{nodes}} \Xi_{Ni}}{n_{graphs}\cdot n_{nodes}}.\label{eq-avgaccposerr}
\end{align}

\subsection{Networks and Training}\label{sec-res-networks-training}
Multiple networks are trained for each of the datasets. These networks are either (relatively) small networks (denoted by the prefix S-) or large networks (denoted by the prefix L-). The size of the small (and large) networks is different for each of the problems and is a function of the hyperparameters. The smallest network has around 600k trainable parameters. The networks are trained in two modes. In `dynamic-only' mode (denoted by suffix -D), the training is only dependent on the dynamic loss, i.e., $(w_d, w_c)=(1,0)$ in Equation \eqref{eq-totalloss}. In `dynamic and contact' mode (denoted by suffix -DC), training starts with a `dynamic-only ' mode for 1000 epochs followed by the inclusion of contact loss afterward, i.e., $(w_d, w_c)\neq(0,0)$ after the thousandth epoch. We do so to avoid an overemphasis on contact loss when the predictions from the surrogate model are bad in its initial training stage. We highlight that the inclusion of contact loss increases the training time significantly, more in Section \ref{sec-train-cost-trade-off}. For this reason, it was not possible for us to perform ablation studies. However, since the S-D and S-DC networks are initialized identically, we believe that a fair comparison is possible.

\begin{figure}[!htb]%% placement specifier
	%% Use \includegraphics command to insert graphic files. Place graphics files in 
	%% working directory.
	\centering%% For centre alignment of image.
	\includegraphics[width=\textwidth]{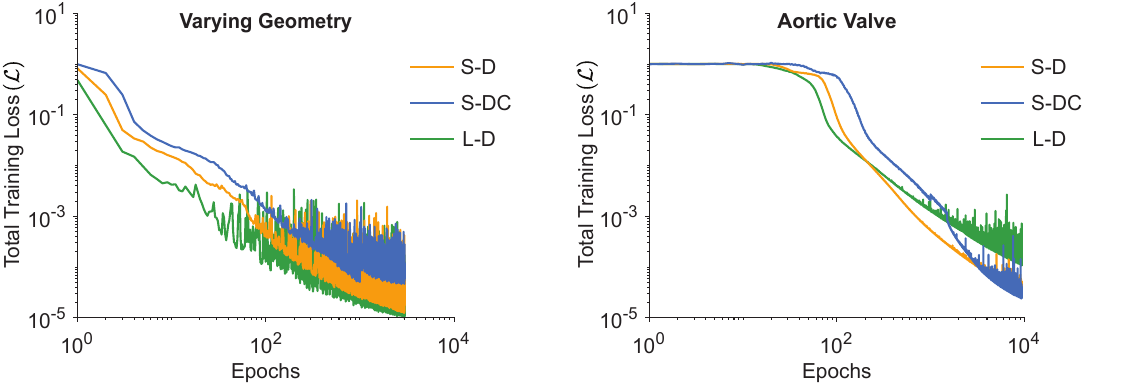}
	%% Use \caption command for figure caption and label.
	\caption{Total $(\mathcal{L})$ over training set for \textbf{(a)} varying geometry and \textbf{(b)} aortic valve problems. Note that the losses are average across all batches at the respective epoch.}\label{fig_res_paperloss}
	%% https://en.wikibooks.org/wiki/LaTeX/Importing_Graphics#Importing_external_graphics
\end{figure}

The weight for $w_d$ is set to unity, and $w_c$ is chosen to bring the computed contact loss to the same scale as the dynamic loss at the thousandth epoch. In the following sections, we look at three networks based on the combination of size and training modes: (S-D, S-DC, L-D). The overall training process is shown in Figure \ref{fig_res_paperloss}, where the total loss ($\mathcal{L}$, Equation \eqref{eq-totalloss}) reduces several orders of magnitude. All hyperparameters of the networks, learning rate schedules, and evolution of each type of loss for all networks and problems are presented in detail in \ref{app-training}. 

\subsection{Impact of Contact Loss Inclusion on Learning}

\begin{figure}[!htb]%% placement specifier
	%% Use \includegraphics command to insert graphic files. Place graphics files in 
	%% working directory.
	\centering%% For centre alignment of image.
	\includegraphics[width=\textwidth]{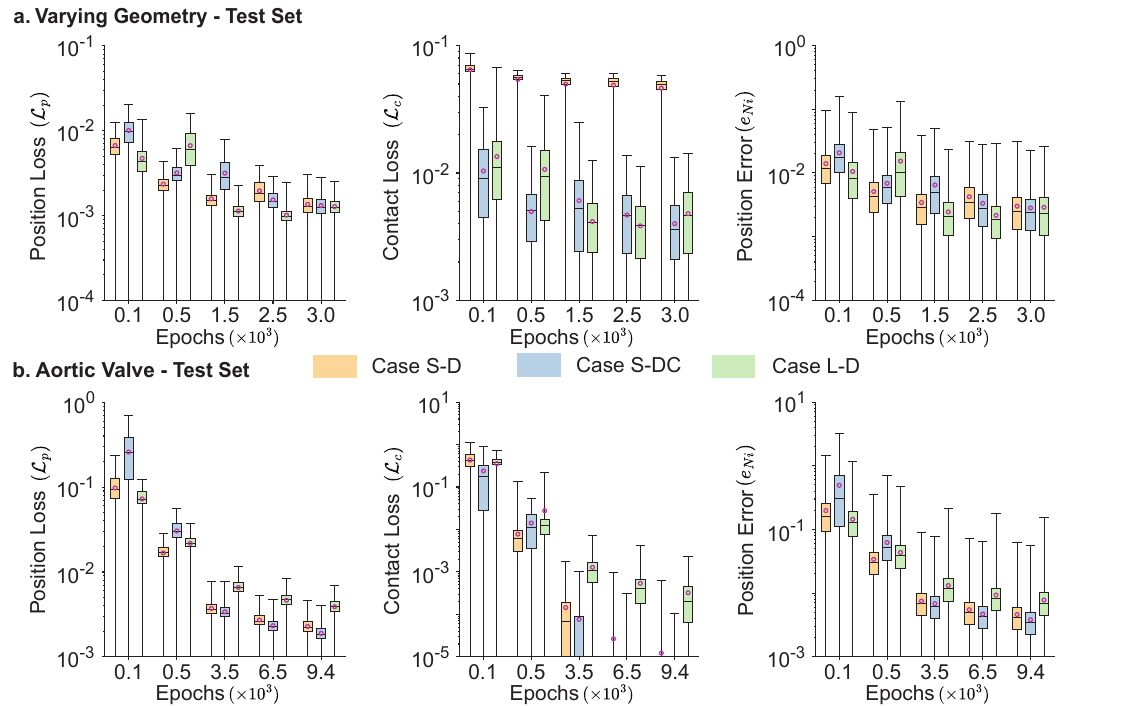}
	%% Use \caption command for figure caption and label.
	\caption{Quartile boundaries (box and whisker) and mean values (circles) of position loss, contact loss, and position error when inference is done on test set at specified epochs during the training process. \textbf{(a)} is for varying geometry problem and \textbf{(b)} for aortic valve problem. Note that for position and contact loss, we evaluate over one graph each in batch dimension axes.}\label{fig_res_multiepoch}
	%% https://en.wikibooks.org/wiki/LaTeX/Importing_Graphics#Importing_external_graphics
\end{figure}

\begin{figure}[!htb]%% placement specifier
	%% Use \includegraphics command to insert graphic files. Place graphics files in 
	%% working directory.
	\centering%% For centre alignment of image.
	\includegraphics[width=\textwidth]{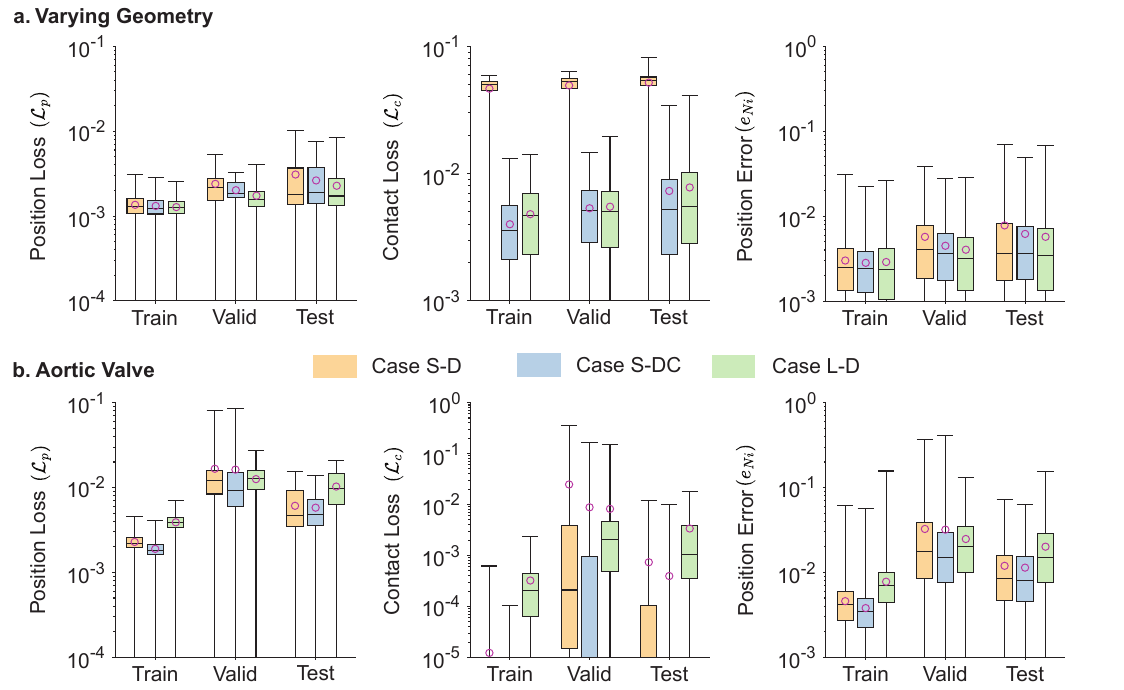}
	%% Use \caption command for figure caption and label.
	\caption{Quartile boundaries (box and whisker) and mean values (circles) of position loss, contact loss, and position error when inference is done using the final trained model on each of training, validation, and test sets. \textbf{(a)} is for varying geometry problem and \textbf{(b)} for aortic valve problem. Note that for position and contact loss, we evaluate over one graph each in batch dimension axes.}\label{fig_res_statsForTrainedModel}
	%% https://en.wikibooks.org/wiki/LaTeX/Importing_Graphics#Importing_external_graphics
\end{figure}

\subsubsection{During the Training Process}\label{sec-during-the-training-process}

During the training process, the loss quantities at each epoch are available as an average of all batches in an epoch. To get finer insights, we use the checkpointed model at regular intervals during the training process and infer on training, validation, and test sets. This inference leads to one quantity per graph for position ($\mathcal{L}_p$, Equation \eqref{eq-position-loss}) and contact ($\mathcal{L}_c$, Equation \eqref{eq-define-contactloss}) loss. We obtain a vector of dimension $n_{nodes}$ for position error ($e_{Ni}$, Equation \eqref{eq-error-set}) at each graph. Figure \ref{fig_res_multiepoch} presents the quartile boundaries and averages of the these quantities at various epochs on the test set. Inference is done at two common epochs (100 \& 500) for both problems, followed by regular sampling until the end of the training.

\paragraph{Aortic Valve (see Figure \ref{fig_res_multiepoch}.b)}  The contact losses for the S-DC network are similar to the S-D and L-D networks before the thousandth epoch, at which point it becomes part of the total training losses. Thereafter, the distribution of the contact loss shifts considerably to lower values for S-DC compared to the S-D  and L-D networks. A similar trend is observed for the position loss and position error; however, unlike contact losses, the position losses and position error start out higher for the S-DC network than for the S-D or L-D networks before overtaking them.

\paragraph{Varying Geometry (see Figure \ref{fig_res_multiepoch}.a)} The reduction in contact and position losses occur as they did in the previous problem. However, both quantities do not undergo an as pronounced reduction as compared to the aortic valve problem, this can be seen by the differences in the scale of the vertical axes. 

\paragraph{Common} Thus, across both benchmark problems, we note that adding a contact loss term aids in learning the contact behavior. Even though the improvements in the positional losses (and error) are marginal, we observe a substantial downward shift of the distribution (and maximum values) for contact loss. \textit{This indicates that the inclusion of contact loss has a regularizing effect as the network progressively trains.}

\subsubsection{On the Final Trained Model}

To see the regularization and generalization effect more clearly, the same metrics are shown for the final epoch and across all dataset splits in Figure \ref{fig_res_statsForTrainedModel}. 
\paragraph{Aortic Valve  (see Figure \ref{fig_res_statsForTrainedModel}.b)} For contact losses, the network S-D performs well on the training set. However, it does not generalize well to either the validation or the test set. This is seen as a shift of the contact loss distribution towards much higher values. While the network S-DC also shifts towards higher contact loss, the shift is not nearly as substantial as for the S-D case.  A much larger network L-D shows similar trends as S-D. The increase in size of the L-D network is not sufficient to counter the improvements brought in by the inclusion of contact loss. \textit{The inclusion of contact loss leads to better generalization in this case.} 
\paragraph{Varying Geometry  (see Figure \ref{fig_res_statsForTrainedModel}.a)} Similar performance (as for the aortic valve) is observed across training, validation, and test sets for each network. The contact losses are much lower for the bulk of the distribution (in the interquartile range) for the S-DC networks than for the S-D network across all dataset splits. Similar to the aortic valve problem, the performance on position error for S-DC is only marginally better than S-D and L-D. However, unlike the previous problem, we note that increasing the size of the network (L-D) without the inclusion of contact losses also leads to similar, but not better, contact loss performance. Note that this is a case of complex contact, specifically, (i) the surfaces do not meet at the same plane in each simulation or the same plane within the same simulation, and (ii) the contact angles of the elements that form the contact pair are not uniform. \textit{The regularization effect and generalization, in a case of complex contact scenario, is observed to be less pronounced.}

\begin{figure}[!htb]%% placement specifier
	%% Use \includegraphics command to insert graphic files. Place graphics files in 
	%% working directory.
	\centering%% For centre alignment of image.
	\includegraphics[width=\textwidth]{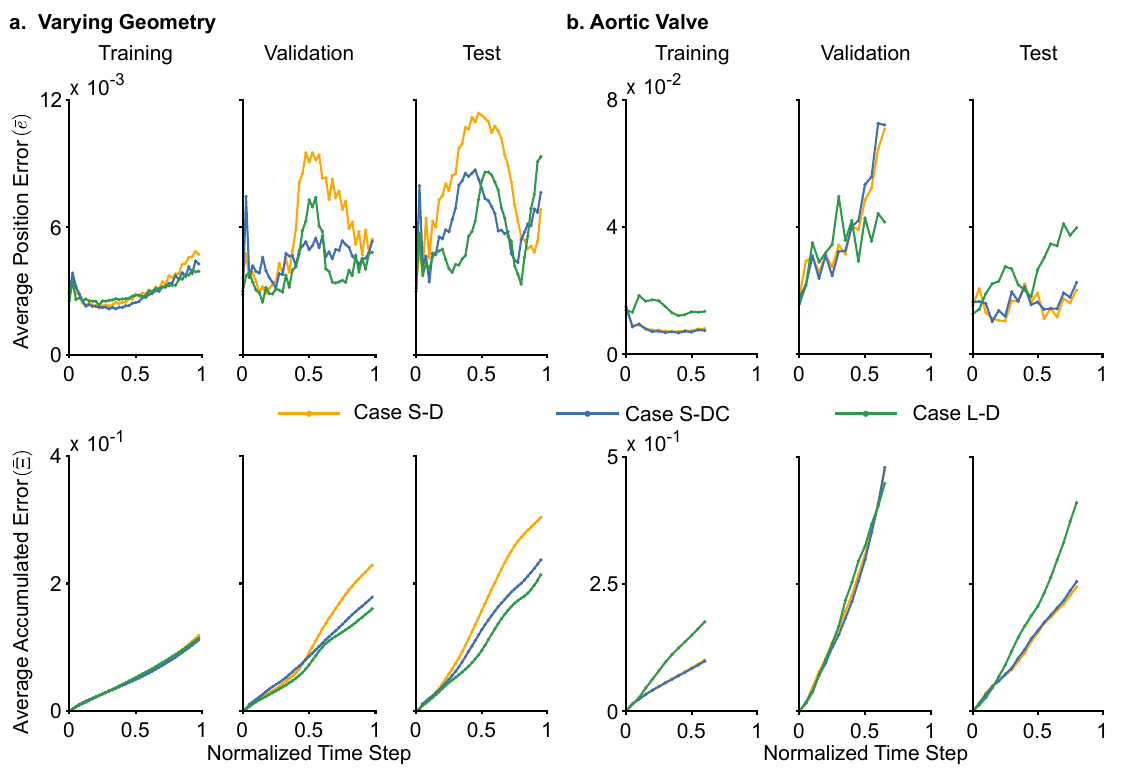}
	%% Use \caption command for figure caption and label.
	\caption{Average position error (Equation \eqref{eq-avgposerr}) and average accumulated position error (Equation \eqref{eq-avgaccposerr}), both averaged across nodes and graphs of simulation at same time step. \textbf{a: Varying Geometry} for first three column, left to right are evaluations over training, validation, and test set. \textbf{b: Aortic Valve} for the last three columns, left to right are evaluations over training, validation, and test set.}\label{fig_res_temporalinstability}
	%% https://en.wikibooks.org/wiki/LaTeX/Importing_Graphics#Importing_external_graphics
\end{figure}

\subsection{Temporal Performance and Instabilities}

\begin{figure}[!htb]%% placement specifier
	%% Use \includegraphics command to insert graphic files. Place graphics files in 
	%% working directory.
	\centering%% For centre alignment of image.
	\includegraphics[width=\textwidth]{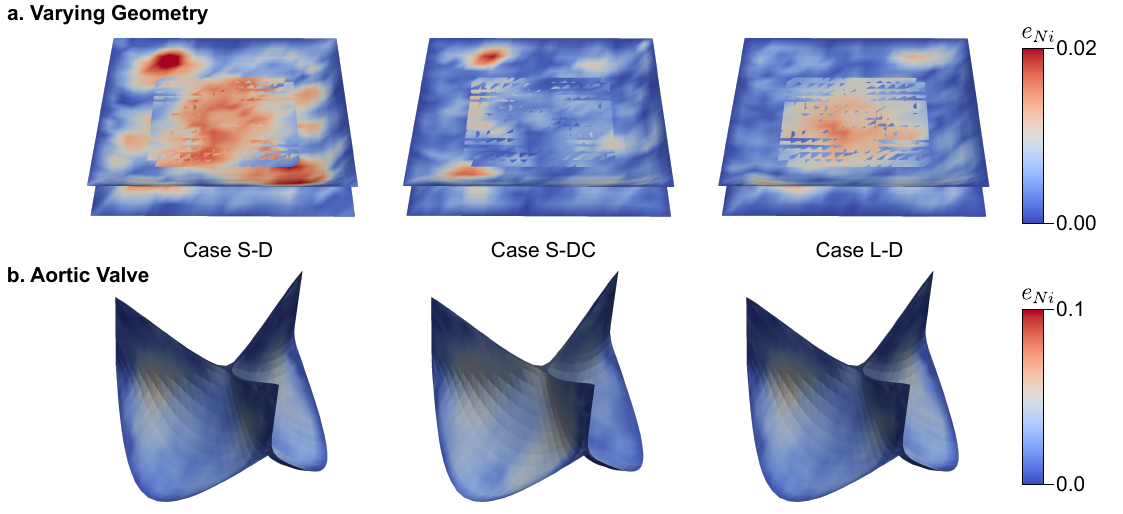}
	%% Use \caption command for figure caption and label.
	\caption{Position error (Equation \eqref{eq-error-set}) for a randomly selected simulation from the test set for each of the problems at the penultimate time step.}\label{fig_res_av_vg_visualization}
	%% https://en.wikibooks.org/wiki/LaTeX/Importing_Graphics#Importing_external_graphics
\end{figure}

The final trained model is used to infer the position errors ($e_{Ni}$, Equation \eqref{eq-error-set}) and accumulated position errors ($\Xi_{Mi}$, Equation \eqref{eq-accposerr}) across all the dataset splits. These quantities are scalars for each node of each graph. We compute the averages over each node of the graph and over all graphs at the same time step across all simulations in training, validation, and test sets. This leads to average position errors ($\bar{e}$, Equation \eqref{eq-avgposerr}) and average accumulated position errors ($\bar{\Xi}$, Equation \eqref{eq-avgaccposerr}). These metrics are plotted against the time step number normalized by the total number of time steps for both problems in Figure \ref{fig_res_temporalinstability}. Note that not all simulations contain an equal number of time steps, and we restrict the plot to the least common time slice across all simulations of the given dataset to allow for meaningful comparison. We also show the position error for each problem and network at one of the timesteps in Figure \ref{fig_res_av_vg_visualization}.

\paragraph{Average Position Error (see top row of Figure \ref{fig_res_temporalinstability})} At each time step, we note that S-D and S-DC networks result in similar average position error $\bar{e}$ on the training set. For the aortic valve problem, this trend extends to validation and test sets. However, for the varying geometry problem, we see that the S-DC network leads to lower $\bar{e}$ than the S-D network during the majority of the time steps. The L-D network performs similarly to S-DC at each time step of the varying geometry problem but does worse for the aortic valve problem. These observations are consistent with the average values of the position error ($e_{Ni}$) seen for all the networks and sets in Figure \ref{fig_res_statsForTrainedModel}. \textit{Although the S-DC network shows marginal improvements in positional error, we see it generalizes better than network without contact (S-D), at least for the varying geometry case.}
\paragraph{Average Accumulated Position Error (see bottom row of Figure \ref{fig_res_temporalinstability})} This measure computes the accumulated position error norms over time. Thus, it establishes an upper bound on the position error if all displacements were summed to obtain the current position. We see an increase in this upper bound as time progresses, for all the networks. \textit{We recognize that this error must be reduced to make the current framework a viable alternative surrogate for nonlinear contact mechanics problems with contact present.} Rolling out, i.e., using the predicted current step for the next step, leads to instabilities. This is expected and detailed in the previous studies using graph neural networks \cite{sanchez-gonzalez_learning_2020}\cite{dalton_emulation_2022}\cite{pfaff_learning_2020}. So far, we have not utilized strategies, such as training with injected noise, to make our networks more robust to these instabilities due to the computational cost (see Section \ref{sec-train-cost-trade-off}), but this remains a future direction.

\subsection{Speedups}
The FEM software FEBio \cite{maas_febio_2012} that has been used to generate the dataset cannot utilize GPU and relies on CPU parallelization. On the other hand, machine learning models can run with CPU but can also take advantage of GPU acceleration. To (i) avoid unfair comparison, and (ii) provide realistic speedups with each hardware accessible to each framework, we provide the run times (inference times) and speedups for three platforms. Platform 1 relies exclusively on CPU and is the only one accessible to the FEM process. Platforms 2 and 3 are for GPU inference, with the former on an NVIDIA Quadro P4000 GPU and the latter on an NVIDIA A100 GPU. Table \ref{tab-inference-times} summarizes the average time per simulation and speedups, all with respect to FEM run times on platform 1. Note that a sixfold speedup is achieved even with just the CPU and a problem as complex as the varying geometry. However, the inferences on platforms 2 and 3 provide over a thousand-fold speedups. On platform 3, five of the six networks provide an inference per simulation within a second.

% \usepackage{multirow}
\begin{table}[hbt]
\normalsize
\centering
\begin{tabular}{c|c|c|cc|cc|cc}
\toprule
\textbf{} & \textbf{} & \textbf{FEM} & \multicolumn{6}{c}{\textbf{Inference}}  \\ 
\textbf{Problem} & \textbf{Network} & Platform 1 & \multicolumn{2}{c}{Platform 1} & \multicolumn{2}{c}{Platform 2} & \multicolumn{2}{c}{Platform 3} \\
\textbf{} & \textbf{} & Time [s] & Time [s] & Speedup & Time [s] & Speedup & Time [s] & Speedup\\ \midrule
\multirow{3}{*}{\begin{tabular}[c]{@{}c@{}}Varying \\ Geometry\end{tabular}} & S-D   & \multirow{3}{*}{1583.7} & 246.80 & 6.4 & 1.36 & 1164.5 & 1.00 & 1583.7  \\
                                                                             & S-DC &                          & 242.84 & 6.5 & 1.36 & 1164.5 & 0.96 & 1649.7 \\
                                                                             & L-D &                       & 794.48 & 2.0 & 3.60 & 439.9 & 1.84 & 860.7\\ \midrule
\multirow{3}{*}{\begin{tabular}[c]{@{}c@{}}Aortic \\ Valve\end{tabular}} & S-D & \multirow{3}{*}{954.1} & 19.48 & 49.0 & 0.40 & 2385.3 & 0.64 & 1490.8 \\
                                                                         & S-DC &                         & 19.24 & 49.6 & 0.44 & 2168.4 & 0.60 & 1590.2 \\
                                                                         & L-D &                      & 98.48 & 9.7 & 0.88 & 1084.2 & 0.88 & 1084.2 \\ \bottomrule
\end{tabular}
\caption{Inference time per simulation in seconds for three platforms and speedups achieved with respect to FEM run on platform 1 for both benchmark problems and all networks. \textbf{Platform 1:} Inference using CPU; Intel(R) Xeon(R) CPU E5-2697 v4 @ 2.30GHz with 16 cores at 1 thread/core and 128GB RAM. \textbf{Platform 2:} Inference using NVIDIA Quadro\textsuperscript{®} Pascal P4000 GPU with memory of 8GB. \textbf{Platform 3:} Inference using NVIDIA A100\textsuperscript{®}GPU with memory of 40GB.}
\label{tab-inference-times}
\end{table}

% \usepackage{multirow}
\begin{table}[]
\centering
\begin{tabular}{c|c|c|c|c|c}
\toprule
\multirow{3}{*}{\textbf{Problem}}           &         & \multicolumn{4}{c}{\textbf{Training Time on Platform 3}}                  \\ 
                                   & \textbf{Network} & 1 epoch       & 1000 epoch  & Total Epochs & Total Training Time \\
                                   &         & {[}in min.{]} & {[}in hr{]} & {[}-{]}      & {[}in days{]}       \\ \midrule
\multirow{3}{*}{Varying  Geometry} & S-D     & 0.21          & 3.48        & 3000         & 0.44                \\
                                   & S-DC    & 8.81          & 146.68      & 3000         & 12.66*               \\
                                   & L-D     & 0.61          & 10.32       & 3000         & 1.29                \\ \midrule
\multirow{3}{*}{Aortic Valve}      & S-D     & 0.10          & 1.74        & 9400         & 0.68                \\
                                   & S-DC    & 2.99          & 49.84       & 9400         & 18.13*               \\
                                   & L-D     & 0.34          & 5.66        & 9400         & 2.22               \\\bottomrule
\end{tabular}
\caption{Total training times for each of the networks for the specified number of epochs, as well as total training times. *: for the S-DC network the first thousand epochs are as fast as the S-D network and differences occur only thereafter; in the table above, times for 1 and 1000 epoch are with contact present, and the total training time takes into consideration that first 1000 epochs happen without contact loss computation. Note that the S-DC networks are the rate-determining network if comparison is to be made for `equally-trained models'. This allowed us to run S-D and L-D networks on slower NVIDIA A40 GPUs to allow more economical use of our computational budget.}
\label{tab-training-times}
\end{table}

\subsection{Training Costs and Trade-offs}\label{sec-train-cost-trade-off}
The Table \ref{tab-training-times} shows the time it takes to train one epoch of a given network on platform 3. For these time profiling, the batch sizes are chosen to make the most efficient use of the GPU memory and vary across the networks. Clearly, the inclusion of contact losses incurs significantly more computational costs (for network S-DC) than (i) training a similar-sized network without contact (S-D) or (ii) a larger network (L-D) with the number of trainable parameters at least three-fold higher than that in the small networks (S-D and S-DC, see Table \ref{tab-architecture-detail}). There is a trade-off between the cost incurred in training and the benefits provided by the regularizing effect (and better generalization) owing to the inclusion of contact loss. Another consideration is the importance of adherence to contact constraints or violations thereof in applications where the current framework might be useful. More specifically, similar adherence when inferring from unseen data. We might be able to swing this trade-off more in favor of the presented framework by improving the computational efficiency of the contact algorithm, possibly through the following directions:
\begin{itemize}
  \item Just-in-time compilation: Currently, although some parts of the contact algorithm are compilable, a wider compilation is not achieved due to constraints that come with such compilation. Reworking the code to make this possible would allow for larger code compilation.
  \item Replacement of current subalgorithms with more efficient ones: (i) for solving linear inequality systems, (ii) for roots of cubic polynomials within a given interval robust enough to handle all root scenarios, and (iii) for axes-aligned bounding boxes, preferably hierarchical and reliant on tree structure.
  \item Reduction in decision-making statements and GPU memory footprint.
  \item Incorporation of multi-GPU distributed training.
\end{itemize}
In shared code base we provide detailed documentation with ample inline comments to make it easy for any reader interested in improving on the current framework. These avenues remain options for the future direction of our work.

\section{Conclusions}\label{sec-conclusion}
% We presented an extension of the graph neural network framework for data-driven learning of displacements for mechanics problems involving contact between two deformable bodies (unlike rigid-soft object interactions). A contact algorithm is implemented that penalizes the contact violation that is (i) of continuous nature, i.e. detects interpenetrations at any point between the time steps, and (ii) relies not just on necessary but also sufficient collision conditions. This framework will be helpful in applications that require fast machine learning-based surrogates for nonlinear mechanics with contact between soft bodies. The implemented framework, at least on the benchmark problems considered, resulted in a regularizing effect and led to better generalization. Moreover, this held for simple contact (contact at similar planes and element normals) and complex contact (contact at differing planes and element normal angles). It should also be noted that for the considered varying geometry example, the current framework had no information on the shape parameters, and yet the benefits were observed despite the varying reference configurations in each simulation. These benefits come in addition to up to a thousand-fold speedups in inference but also come at tremendous computational costs of training.

We presented an extension of the data-driven graph neural network to learn contact between two deformable bodies (unlike rigid-soft interactions). A contact algorithm is implemented that relies not only on necessary but also on sufficient collision conditions. This algorithm is continuous; i.e., it detects interpenetrations at any point between the time steps. This enables calculation of the contact loss term, which penalizes contact violations during training. This framework will be particularly helpful in applications that require rapid machine learning-based surrogates for nonlinear mechanics involving contact between soft bodies. The implemented framework, at least on the benchmark problems considered, resulted in a regularizing effect and led to better generalization. Moreover, this held for simple contact (contact at similar planes and element normals) and complex contact (contact at differing planes and element normal angles). It should also be noted that, for the considered varying geometry benchmark problem, the current framework had no information on the shape parameters. Yet, benefits were observed despite the varying reference configurations in each simulation. These benefits come in addition to up to a thousand-fold speedups in inference, but also incur tremendous computational costs during training.

%Overall, we believe that the findings can help the community in making steps toward improving surrogate models for contact mechanics applications. We believe that the idea of improving graph neural networks by using robust contact detection algorithms and then penalizing contact is obvious and might have been tried earlier if these algorithms were easy to implement. Even though, it seems that the computational cost of contact detection in an automatic differentiation framework with a large graph/mesh is (almost) prohibitively expensive, we believe that our work offers the community a practical benchmark for weighing its benefits against its costs.

The idea of improving graph neural networks by utilizing robust contact detection algorithms and then penalizing contact violations is straightforward. However, this direction is relatively unexplored, likely due to the considerable efforts required to implement a comprehensive contact detection algorithm within an automatic differentiation framework with a large graph. We find that although there are benefits to moving in this direction, it comes at a computational cost that can be prohibitive. Through this work, we share with the community a practical benchmark for weighing its benefits against its costs. 

\section*{Code and Data Availability}\label{sec-codebase-data}
%All the codes are available at \href{?}{Github} and the dataset is available at \href{?}{Texas DataVerse}. \textcolor{red}{Links would be updated before publication to correct pointers.}.

After the acceptance of this manuscript, all the code and data will be made available through \href{https://github.com/SoftTissueBiomechanicsLab/GraphNeuralNetwork_SoftBodyContact}{this github repository} and \href{https://doi.org/10.18738/T8/HTGUP6}{this DataVerse} repository, respectively.

\section*{Acknowledgments}

This work was funded by NIH R01HL165251, NIH R21HL161832, U.S. NSF CMMI awards 2438943, 2235856, and 2127925 (to MKR). We acknowledge additional computing support from the Texas Advanced Computing Center (Project: DMS24010). This work also used the Delta system at the National Center for Supercomputing Applications through allocation CIS240404 from the Advanced Cyberinfrastructure Coordination Ecosystem Services \& Support (ACCESS) program. We also thank undergraduate researcher Manas Pathak for their efforts in debugging parts of the contact algorithm implementation.

\appendix

\section{Training Process}\label{app-training}
\begin{figure}[!htb]%% placement specifier
	%% Use \includegraphics command to insert graphic files. Place graphics files in 
	%% working directory.
	\centering%% For centre alignment of image.
	\includegraphics[width=\textwidth]{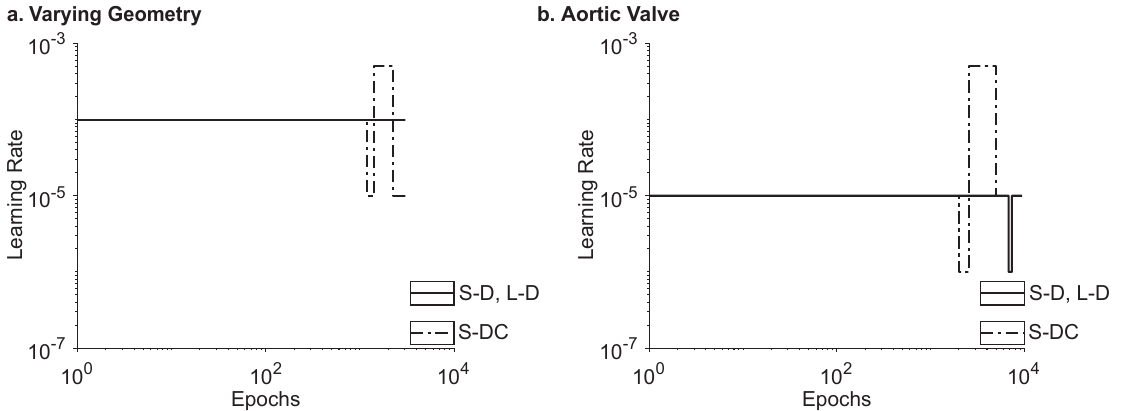}
	%% Use \caption command for figure caption and label.
	\caption{Learning rate schedule for all networks for both the problems.}\label{fig_res_learningrate}
	%% https://en.wikibooks.org/wiki/LaTeX/Importing_Graphics#Importing_external_graphics
\end{figure}

The step function learning rate schedule is used to train each network and is shown in the Figure \ref{fig_res_learningrate}. The learning rate schedules are identical for the `-D' mode training and slightly different (near the end) for the `-DC' training mode. The extensive exploration of the hyperparameter space or more experiments with different learning rate schedules could not be carried out due to the high computational costs of the training. This training cost is higher for the varying geometry problem than for the aortic valve problem due to the larger dataset size. The training of networks in `-D' is computationally cheaper. Still, we select the equally trained model at the last common checkpointed epoch to present a fair comparison across all networks. Due to this consideration, the varying geometry and aortic valve problems are trained to 3000 and 9400 epochs, respectively. For the optimizer, we utilize weight decay during the learning process. Numerical stability is improved by: (i) using available options in the Adam optimizer implementation in PyTorch, i.e. setting the $\epsilon$ parameter, and (ii) enabling gradient norm clipping.

Options to improve numerical stability available in Adam optimizer implementation of PyTorch are also used in addition to gradient norm clipping.

The network parameters for each encoder, processor, and decoder MLPs are shown in the Table \ref{tab-architecture-detail}. We also report the dimension of each feature space and its corresponding embedding space. Note that different MLPs of the same size are used for each round of message passing. The rounds of messaging passing for each network are reported, too. Finally, we compute the total number of trainable parameters for each of the networks and report those. 
% \usepackage{booktabs}
% \usepackage{multirow}
\begin{table}[!hbt]
\centering
\normalsize
\begin{tabular}{@{}c|cc|cc@{}}
\toprule
\multirow[c]{2}{*}{\textbf{\begin{tabular}[c]{@{}c@{}}\\ Parameter\end{tabular}}} &
  \multicolumn{2}{c}{\textbf{Varying Geometry}} &
  \multicolumn{2}{c}{\textbf{Aortic Valve}} \\ 
 &
  \textbf{\begin{tabular}[c]{@{}c@{}}\midrule Small Network \\ S-D/S-DC\end{tabular}} &
  \textbf{\begin{tabular}[c]{@{}c@{}}\midrule Large Network\\ L-D\end{tabular}} &
  \textbf{\begin{tabular}[c]{@{}c@{}}\midrule Small Network\\ S-D/S-DC\end{tabular}} &
  \textbf{\begin{tabular}[c]{@{}c@{}}\midrule Large Network\\ L-D\end{tabular}} \\
  \midrule
$\mathfrak{D}(x_{Ii})$                   & 9          & 9           & 9          & 9           \\
$\mathfrak{D}(e^M_{Iij})$                & 4          & 4           & 4          & 4           \\
$\mathfrak{D}(e^W_{Iij})$                & 4          & 4           & 4          & 4           \\
$\mathfrak{D}(\mathbf{x}_{Ii})$          & 64         & 64          & 64         & 64          \\
$\mathcal{W}(\mathcal{E}_x)$             & 128        & 128         & 128        & 128         \\
 $\mathbb{D}(\mathcal{E}_x)$             & 3          & 3           & 2 & 4*  \\
$\mathfrak{D}(\mathbf{e}^M_{Iij})$       & 64         & 64          & 64         & 64          \\
$\mathcal{W}(\mathcal{E}_e^M)$           & 128        & 128         & 128        & 128         \\
 $\mathbb{D}(\mathcal{E}_e^M)$           & 3          & 3           & 2 & 4*  \\
$\mathfrak{D}(\mathbf{e}^W_{Iij})$       & 64         & 64          & 64         & 64          \\
$\mathcal{W}(\mathcal{E}_e^W)$           & 128        & 128         & 128        & 128         \\
 $\mathbb{D}(\mathcal{E}_e^W)$           & 3          & 3           & 2 & 4*  \\
$\mathcal{W}(\phi^{M(n)})$               & 128        & 128         & 128        & 128         \\
 $\mathbb{D}(\phi^{M(n)})$               & 3          & 3           & 2          & 2           \\
$\mathcal{W}(\phi^{W(n)})$               & 128        & 128         & 128        & 128         \\
 $\mathbb{D}(\phi^{W(n)})$               & 3          & 3           & 2          & 2           \\
 $\mathbb{D}(\gamma^{(n)})$              & 128        & 128         & 128        & 128         \\
 $\mathbb{D}(\gamma^{(n)})$              & 3          & 3           & 2          & 2           \\
$k$                                      & 5          & 20* & 3 & 20* \\
$\mathcal{W}(\mathcal{D}_i)$             & 128        & 128         & 128        & 128         \\
 $\mathbb{D}(\mathcal{D}_i)$             & 2 & 3*  & 2 & 4*  \\
$\mathfrak{D}(\hat{y}_{Ii})$             & 3          & 3           & 3          & 3           \\
$\mathfrak{D}(g_I)$                      & 4          & 4           & 6          & 6           \\
$\mathcal{W}(\mathcal{E}_g)$             & 128        & 128         & 128        & 128         \\
 $\mathbb{D}(\mathcal{E}_g)$             & 2 & 3*  & 2 & 4*  \\
\midrule
\textbf{Trainable Parameters}            &  \textbf{1,216,451} &  \textbf{4,251,779} &  \textbf{622,659} &  \textbf{3,376,899} \\ \bottomrule
\end{tabular}
\caption{Neural network size and total number of trainable parameters for each network. $\mathfrak{D}(\cdot)$ is dimension along non-batch axis, $\mathcal{W}(\cdot)$ is width of MLP hidden layer, and $\mathbb{D}(\cdot)$  is depth of MLP where $(\cdot)$ or other symbols follows the notation in Section \ref{sec-architecture}. Parameters that are changed for the large network compared to the small network are denoted using (.)*.}
\label{tab-architecture-detail}
\end{table}

\begin{figure}[!htb]%% placement specifier
	%% Use \includegraphics command to insert graphic files. Place graphics files in 
	%% working directory.
	\centering%% For centre alignment of image.
	\includegraphics[width=0.75\textwidth]{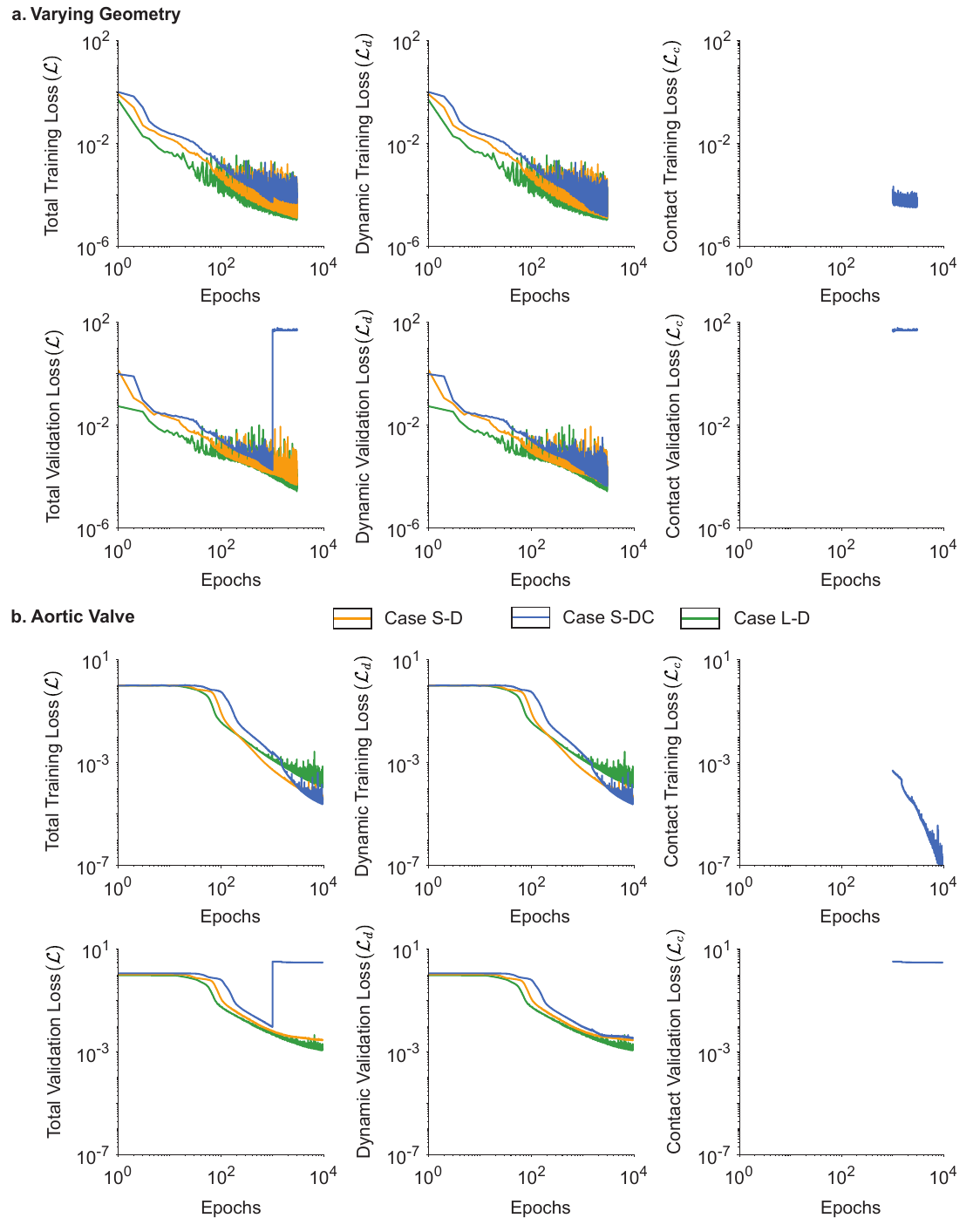}
	%% Use \caption command for figure caption and label.
	\caption{Total $(\mathcal{L})$, contact $(\mathcal{L}_c)$, and dynamic $(\mathcal{L}_d)$ loss over training and validation sets for \textbf{(a)} varying geometry and \textbf{(b)} aortic valve problems. Note that the losses are average across all batches at the respective epoch.}\label{fig_res_appendixloss}
	%% https://en.wikibooks.org/wiki/LaTeX/Importing_Graphics#Importing_external_graphics
\end{figure}

See Figure \ref{fig_res_appendixloss} for the complete evolution of all losses through the training process. Note the characteristic jump in the total loss on the validation set for `-DC' mode training for both problems at the 1000th epoch. This is because the weight $(w_c)$ is chosen using the information in the training set and does not guarantee contact losses on the same scale as dynamic losses for the validation set. Since contact computation is expensive, we do not compute the contact loss for `-D' mode learning. However, inference using checkpointed models at periodic epochs fills in for this information as described in Section \ref{sec-during-the-training-process}.    

\section{Dataset Generation Details}\label{app-datasetdetails}
Each data set is generated using the finite element method, where the solver is FEBio. Each mesh is a four-node quadrilateral shell in the upstream finite-element simulation. However, before being used in the current framework, they are all converted to a three-node triangular shell mesh. This is done using the developed continuous collision detection algorithm.  The number of time steps per simulation is the maximum that occurs across all the simulations. The time step size is the same for all the graphs except the final time step, which is set to zero. The framework learns to predict no displacement for this final step. The collision radius $R$ is used for proximity criteria to build world-space edges. The length scale $l_c$ is used to normalize length-based quantities and is of the order of the geometric dimension in the reference state. 

\begin{figure}[!htb]%% placement specifier
	%% Use \includegraphics command to insert graphic files. Place graphics files in 
	%% working directory.
	\centering%% For centre alignment of image.
	\includegraphics[width=0.8\textwidth]{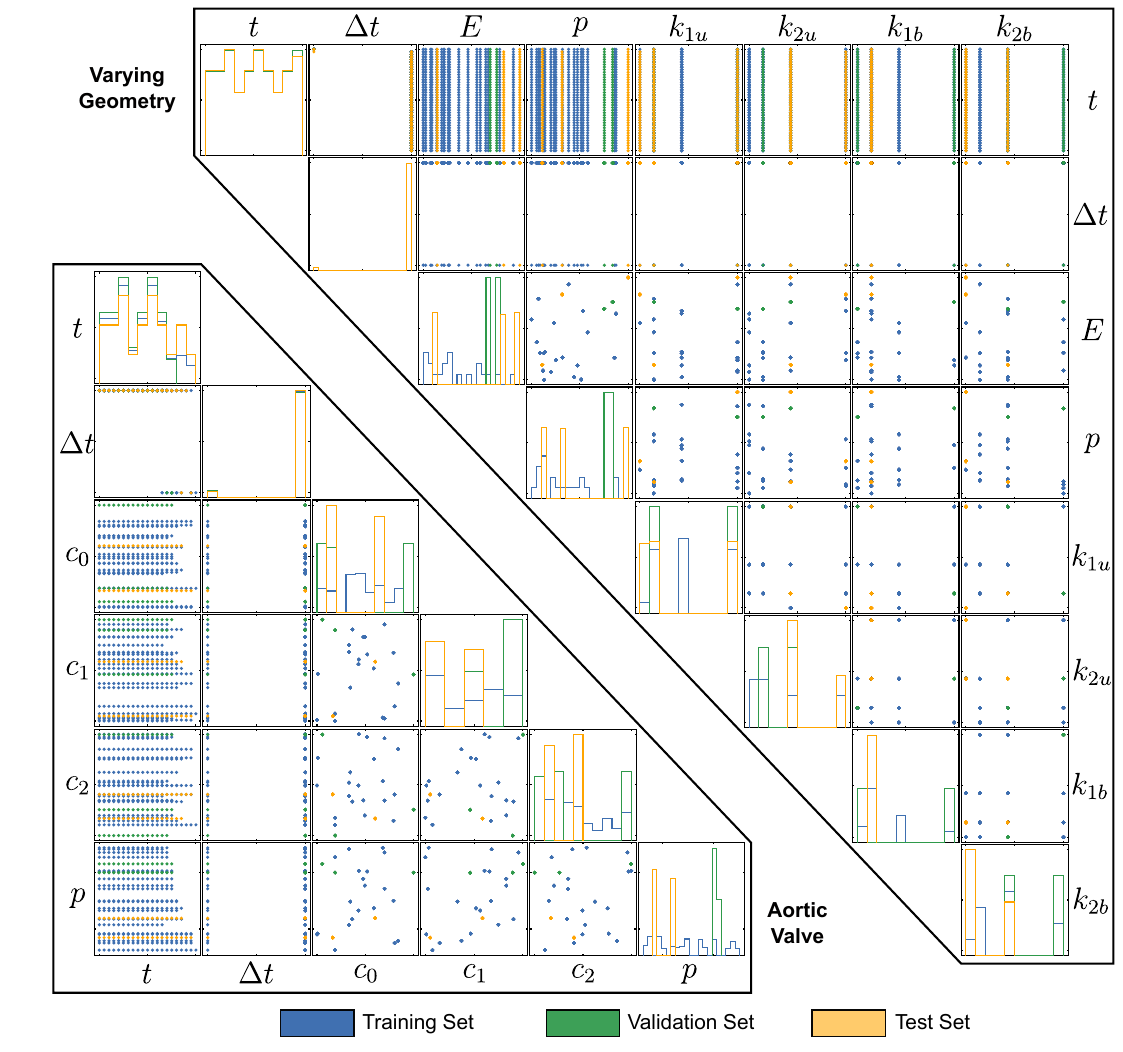}
	%% Use \caption command for figure caption and label.
	\caption{Pairwise scatter plot of all parameters used to generate the dataset labelled by the set they belong to. Also shown are grouped histograms of each parameter. Each parameter has been min-max normalized to $[0,1]$ interval. \textbf{Varying Geometry:} Top right block \textbf{Aortic Valve:} Bottom left block.}\label{fig_graph_features}
	%% https://en.wikibooks.org/wiki/LaTeX/Importing_Graphics#Importing_external_graphics
\end{figure}

Figure \ref{fig_graph_features} shows the scatter plot of the parameters (super-set of graph-level features, mentioned in Section \ref{sec-res-networks-training}) against each other along with their grouped histograms. The group labels are the training, validation, and test sets. 

\bibliographystyle{elsarticle-num} 
\bibliography{GraphNeuralNetworkPaper.bib}

\end{document}